\documentclass[aps,prx,twocolumn,superscriptaddress]{revtex4-2}

\usepackage{amssymb,amsfonts,amsbsy,amsmath}
\usepackage{graphicx}
\usepackage{dcolumn}
\newcolumntype{d}{D{.}{.}{2.1}}
\usepackage{colortbl}
\usepackage[dvipsnames]{xcolor}
\usepackage{upgreek}
\usepackage{bm}
\usepackage[caption=false]{subfig}
\usepackage{tikz}
\usetikzlibrary{shapes,shadows,arrows,positioning,patterns,decorations.pathreplacing,calc,patterns,patterns.meta,snakes,tikzmark}
\usepackage{hyperref}
\usepackage[capitalise]{cleveref}
\usepackage{enumitem}
\setlist[itemize]{leftmargin=*,nosep} 
\setlist[enumerate]{leftmargin=*,nosep} 
\usepackage{nomencl}
\makenomenclature

\begin{document}

\title{Unifying same- and different-material particle charging through stochastic scaling}

\author{Holger Grosshans}
\thanks{Corresponding author}
\email{holger.grosshans@ptb.de}
\affiliation{Physikalisch-Technische Bundesanstalt (PTB), Braunschweig, Germany}%
\affiliation{Otto von Guericke University of Magdeburg, Institute of Apparatus and Environmental Technology, Magdeburg, Germany}
\author{Gizem Ozler}
\affiliation{Physikalisch-Technische Bundesanstalt (PTB), Braunschweig, Germany}%
\affiliation{Otto von Guericke University of Magdeburg, Institute of Apparatus and Environmental Technology, Magdeburg, Germany}
\author{Vyshnavi Veeravalli}
\affiliation{Physikalisch-Technische Bundesanstalt (PTB), Braunschweig, Germany}%
\affiliation{Otto von Guericke University of Magdeburg, Institute of Apparatus and Environmental Technology, Magdeburg, Germany}
\author{Simon Janta\v{c}}
\affiliation{Physikalisch-Technische Bundesanstalt (PTB), Braunschweig, Germany}

\date{\today}

\begin{abstract}
Triboelectric charging of insulating particles through contact is critical in diverse physical and engineering processes, from dust storms and volcanic eruptions to industrial powder handling.
However, many experiments over the years have consistently revealed counterintuitive charging patterns, including variable impact charge under identical conditions, charge sign reversal with repeated impacts, and bipolar charging of differently sized particles.
Existing computational models cannot predict these patterns;
they either rely on oversimplified heuristics or require inaccessible detailed surface properties.
We present a stochastic scaling model (SSM) for particle charging that unifies same-material (particle-particle) and different-material (particle-wall) charging in a single theoretical framework.
The model grounds in a physics-based stochastic closure by the mean, variance, skewness, and minimum impact charge measured in a highly-controlled reference experiment.
To test the SSM, we implemented it in an open-source Lagrangian-Eulerian CFD solver.
When simulating 300\,000 insulating particles transported by turbulent wall-bounded flows, the SSM takes less than 0.01\% of the CPU time.
By scaling the statistical parameters of the reference impact to each collision, the new model reproduces the complex charging patterns observed in experiments without requiring surface-level first-principles inputs.
The SSM offers a physically grounded route to large-scale simulations of electrostatic effects across many fields of particle-laden flows.
\end{abstract}

\maketitle


\section{Introduction}

Contact charging of particles occurs in nature, such as in volcanic plumes~\citep{Cim22}, dust storms on Earth~\citep{Kam72} and other planets~\citep{Shi06}, on asteroids~\citep{Hart19}, and in everyday life, for example, when grinding coffee~\citep{Har24}.
In industrial powder processing, the highest charge typically appears when particles are pneumatically conveyed through ducts~\citep{Kli18}, i.e., when they are transported by a turbulent wall-bounded flow.
In the ducts and downstream components, the charge often causes wall fouling~\citep{Giffin2013} and dust explosions~\citep{Osh11}.

Insulating particles are of particular concern due to their tendency to accumulate and retain high electrostatic charge.
Earth's surface~\citep{Gray24}, coffee grinders~\citep{us10912418}, and industrial transport ducts are typically grounded and conductive, the latter to comply with safety regulations~\citep{60079-32-1}.
Despite the importance of charging insulating particles in turbulent flow along a conductive surface, there is currently no physically grounded charging model that is compatible with computational fluid dynamics~(CFD) simulations and that can reproduce the key experimental observations detailed below:
variable impact charge under identical conditions~\citep{Mat03,Mat06c,Grosj23a}, charge sign reversal with repeated impacts~\citep{Low86b,Shaw28}, and bipolar charging of differently sized particles~\citep{Wait14,Forw09}.

A straightforward modeling approach would involve creating a large database of experimentally measured impact charges under various conditions.
However, this strategy is infeasible in practice.
For a single particle-surface combination, the impact charge depends on numerous variables, including temperature, humidity, pressure, impact mode, angle, velocity, particle size, shape, surface roughness, and initial charge.
Building a statistically relevant dataset would require an impractically large number of experiments.

Even if such a dataset existed, experimental conditions often do not align with those in powder flows.
For instance, the velocity-dependent charging model~\citep{Gro21c} required extrapolation to low impact velocities~($<$5~m/s) that were not accessible in the experimental setup.

The most widely used charging model in CFD is the condenser model, introduced more than five decades ago~\citep{Soo71,Mas76,John80}.
It is derived from electrostatics of conductive materials, where charge transfer is driven by the contact potential difference;
thus, it applies to different-material charging.
The condenser model appeared in many different mathematical formulations (see e.g.,~\citep{Lau13,Mat00,Pei13} or the reviews by~\citep{Gro23d,Mat10,Chow21}).
In the most general form, it gives the impact charge of particle $i$ on a wall of another material by an equation of the type
\begin{equation}
\label{eq:}
\Delta Q_{iw} = h_1(A,\Delta U) - h_2(A,Q) \, .
\end{equation}
That means the impact charge is a function of the contact area, $A$, the contact potential difference between the two materials, $\Delta U$, and the charge the particle holds before impact, $Q$, amongst other electrical and mechanical parameters.

\begin{figure*}[tb]
\begin{center}
\subfloat[Condenser model]{
\begin{tikzpicture}[font=\scriptsize]
\draw [<->] (0,2) node [left] {$Q$} -- (0,0) -- (3.5,0) node [below left] {\# Impact};
\draw [dashed] (0,1.6) node [left] {$Q_\mathrm{sat}$} -- (3.5,1.6);
\node (a) at (0,0) {};
\node (b) at (.6,.8) {};
\node (c) at (1.2,1.2) {};
\node (d) at (1.8,1.4) {};
\node (e) at (2.4,1.5) {};
\node (f) at (3,1.52) {};
\fill[red, radius=2pt] (a) circle [];
\fill[red, radius=2pt] (b) circle [];
\fill[red, radius=2pt] (c) circle [];
\fill[red, radius=2pt] (d) circle [];
\fill[red, radius=2pt] (e) circle [];
\fill[red, radius=2pt] (f) circle [];
\draw (a) -- (b) -- (c) -- (d) -- (e) -- (f);
\draw (b) -- ++(.6,0) -- (c) node [midway,right] {$\Delta Q_{iw}$};
\end{tikzpicture}
\label{fig:condensera}}
~
\subfloat[Variable impact charge~\citep{Grosj23a}]{\includegraphics[trim=0mm 0mm 0mm 210mm,clip=true,width=.28\textwidth]{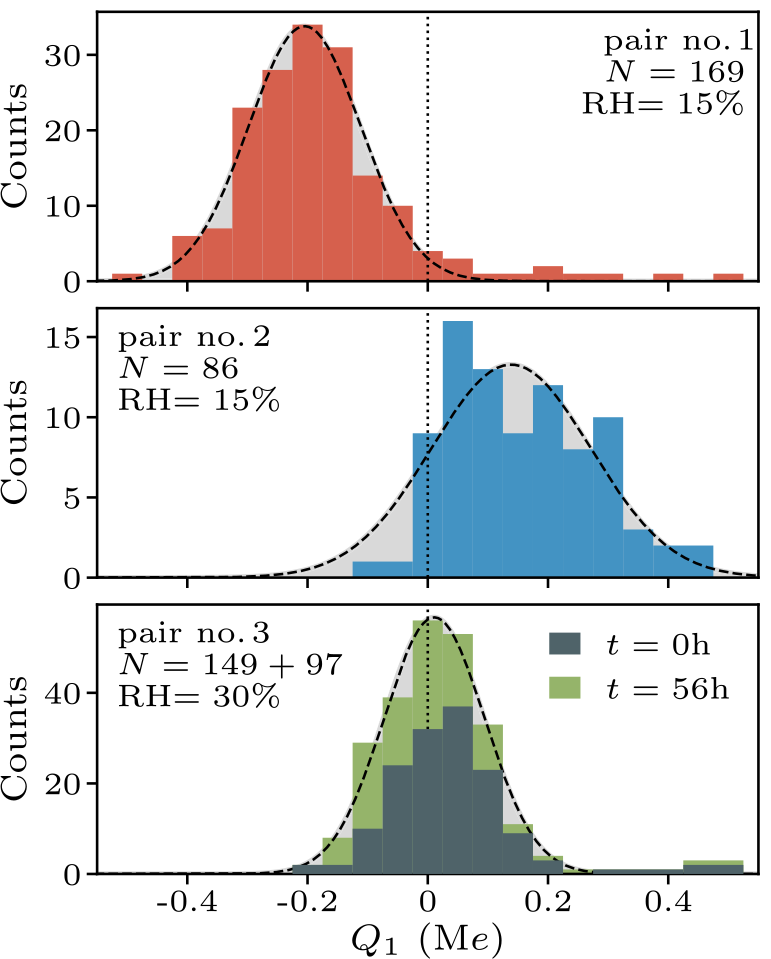}
\label{fig:condenserb}}
~
\subfloat[Charge reversal~\citep{Shaw28}]{
\includegraphics[trim=0mm 70mm 140mm 0mm,clip=true,height=.15\textwidth]{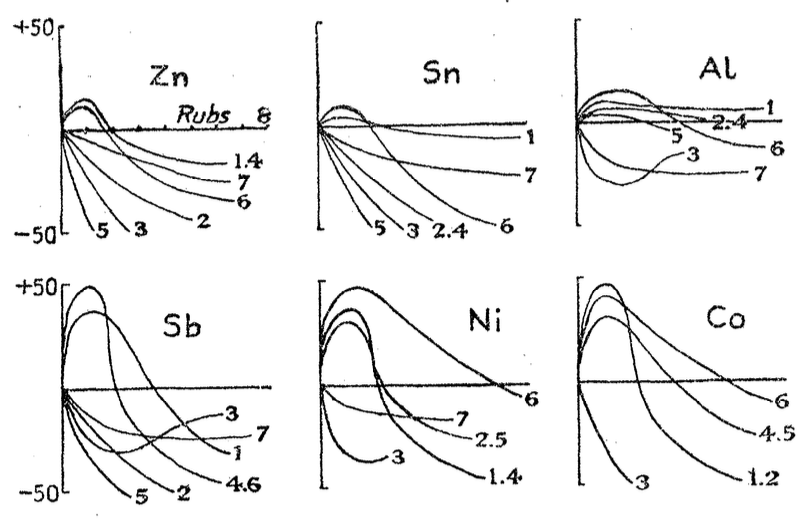}
\label{fig:condenserc}}
~
\subfloat[Bipolar charging~\citep{Forw09}]{
\begin{tikzpicture}[font=\scriptsize]
\node[anchor=south,inner sep=0pt] at (0,0) {\includegraphics[trim=0mm 154mm 30mm 130mm,clip=true,width=.25\textwidth]{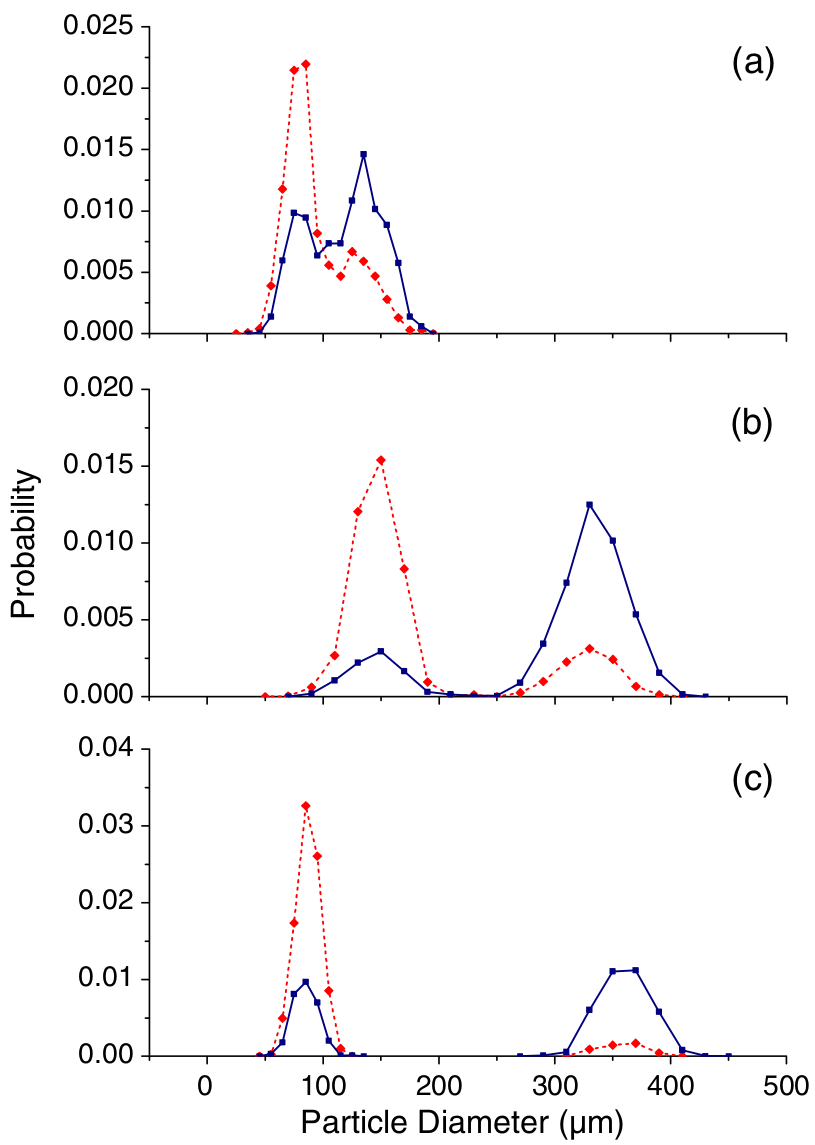}};
\node[anchor=north,inner sep=0pt] at (0,0) {\includegraphics[trim=0mm 0mm 30mm 378mm,clip=true,width=.25\textwidth]{fig/forward}};
\node[red] at (-.1,1.7) {positive};
\node[blue] at (1.4,1.7) {negative};
\end{tikzpicture}
\label{fig:condenserd}}
\end{center}
\caption[]{%
\label{fig:condenser}
(a) The classical condenser model predicts impact charge between different materials ($\Delta Q_{iw}$) according to their contact potential difference, resulting in monotonic charging of a particle until it reaches its saturation charge ($Q_\mathrm{sat}$). 
The model fails to reproduce several experimentally observed charging patterns:
(b)~stochastic variability in impact charge under identical conditions, 
(c)~charge reversal, where the direction of net charge transfer between materials can invert after repeated contacts,
and (d)~size-dependent bipolar charging in contacts between particles of the same material but different sizes
(adapted and reprinted with permission).}
\end{figure*}

The condenser model predicts that a particle repeatedly contacting another material charges until it saturates, i.e., when $h_1$ balances $h_2$ (\cref{fig:condensera}).
The model is fast and its equation system is closed;
its parameters, such as permittivity, surface resistivity, contact potential difference, and effective separation distance, can be measured experimentally.
Also, the condenser model is generally compatible with multiphase flow simulations.
For example, it has been successfully applied in Euler-Lagrange~\citep{Gro16a,Gro17a} and in Euler-Euler~\citep{Gro20h,Ray18,Ray20} formulations.
However, its application to systems involving insulating materials requires extensive parameter tuning and yields limited predictive power.

In particular, the condenser model fails to reproduce several experimentally observed patterns, out of which we highlight three that we consider most important:
\begin{enumerate} 
\item \emph{Variable impact charge:} The condenser model is deterministic, predicting a fixed impact charge for given material properties and impact conditions.
In contrast, same- and different-material experiments with polymer particles show high variability even under carefully controlled conditions (\cref{fig:condenserb})~\citep{Mat03,Mat06c,Grosj23a}.
\item \emph{Charge reversal:} Under repeated contacts of different materials or asymmetric sliding, the direction of net charge transfer can reverse (\cref{fig:condenserc})~\citep{Low86b,Shaw28}, a phenomenon the condenser model cannot predict due to its contact potential difference-driven directionality.
\item \emph{Size-dependent bipolar charging:} Collisions between particles of the same material often result in particles of different sizes acquiring opposite charges (\cref{fig:condenserd})~\citep{Wait14,Forw09}, while the particles' sizes play no role in the condenser model.
\end{enumerate}
All these phenomena were explained in a somewhat similar way, by charging sites~\citep{Bay12,Grosj23a}, films~\citep{Shaw28}, or charge carriers~\citep{Lacks08,Lacks07}, on the material surfaces that behave stochastically or change in time.
Various extensions to the condenser model have been proposed to address specific shortcomings, such as charge scatter~\citep{Gro16f} or size-dependent effects~\citep{Liu20}, but none offers a general or robust solution.

The surface state model, as an alternative to the condenser model, focuses on charging between insulating materials with homogeneous surface composition~\citep{Low86a,Low86b,Lacks08,Lacks07}.
This model introduces the concept of a limited reservoir of transferable charge carriers on each particle's surface, $\Phi$.
The difference of the transferable charge carriers between two particles, $\Delta \Phi$, and the contact area, amongst other parameters, drive the net charge transfer,
\begin{equation}
\label{eq:}
\Delta Q_{ij} = h_3(A,\Delta \Phi) \, .
\end{equation}
As charge transfers, this reservoir depletes.

The surface state model can, contrary to the condenser model, predict size-dependent bipolar charging~\citep{Kon17}.
For wall-bounded turbulent flow, this model showed that large particles charge positively and mid-sized ones negatively when colliding with other particles~\citep{Gro23c}.
However, it applies only to contacts between insulators of the same materials, not to the critical case of particle-wall (insulator-conductor) contacts.
Just like the condenser model, the surface state model cannot predict variable impact charge and charge reversal.
Furthermore, $\Phi$ is generally unknown and treated as a tunable parameter.

At a more fundamental level, mosaic or patch models describe contact charging from first principles by resolving the contacting material's surfaces.
On the resolved surface, they model charge transfer between nanoscopic donor and acceptor sites~\citep{Apo10}.
Charge transfer occurs probabilistically at these sites, resulting in a net stochastic outcome.
This nanoscale mechanism has been qualitatively validated using Kelvin probe force microscopy (KPFM)~\citep{Bay11,Ter89,Burgo12}.

A recent extension of the mosaic model incorporates global differences in donor/acceptor densities~\citep{Grosj23}.
This extension has significant consequences:
Now, the mosaic model unifies same-material and different-material charging, and it captures all three charging patterns listed above.
However, the model's computational complexity limits its applications to domains spanning a few micrometers;
it cannot simulate large-scale particle systems.
Moreover, it depends on inaccessible nanoscopic surface quantities, such as the size of charging sites, site correlation lengths, and transfer probabilities.

To address these limitations, a scaling methodology has recently been developed to extend the mosaic model to full-scale powder flow simulations using discrete element modeling~\citep{Zhang24}.
Yet this approach applies only to same-material charging and requires unmeasured quantities, such as the size of the charging sites, transfer probabilities, and donor densities.

In this work, we propose the stochastic scaling model (SSM) that captures the stochastics and patterns of particle contact charging while remaining computationally efficient and compatible with CFD-scale simulations.
The model builds on the charging physics of the mosaic model by \citet{Grosj23}, but introduces an entirely new set of equations that circumvents the need to resolve the contact surface area.
The SSM generalizes to particle-particle (insulator-insulator) and particle-wall (insulator-conductor) interactions.
It treats the impact charge as a random variable, whose statistical distribution, defined by its mean, standard deviation, and skewness, is derived from particle impact experiments under reference conditions.
These distributions are then scaled based on local impact parameters to generate realistic charge outcomes throughout a simulated powder flow.
The SSM equation system is closed;
it does not rely on any tuning through unmeasurable quantities.

To test and demonstrate the SSM, we implemented it into an Euler-Lagrange solver, even though the SSM (as the condenser model) is generally compatible with multiphase flow simulations. 
We simulate wall-bounded turbulent flows, specifically duct flows such as in pneumatic conveyors, where the particles typically gain the highest charge~\citep{Kli18}.
In our tests, the SSM treats complete powder surface areas that are up to eight orders of magnitude larger than those computed by the mosaic model~\citep{Grosj23}, thereby opening the door to large-scale CFD simulations of electrostatic effects in particle-laden flows.

In \cref{sec:spc}, we derive the SSM formulation.
\Cref{sec:reference-impact} describes the experimental procedure to extract the statistical parameters required as input to the model and to validate the scaling laws.
In \cref{sec:cfd}, we present large-scale CFD simulations that use the SSM to predict the charging of poly(methyl methacrylate) (PMMA) and polystyrene (PS) particles in a canonical turbulent flow bounded by an aluminum surface.
Finally, in \cref{sec:discussion}, we discuss the model’s errors, predictive capabilities, and computational advantages.

\section{Stochastic scaling model (SSM)}
\label{sec:spc}

\subsection{Overview of the model concept}
\label{sec:overview}

\begin{figure*}[tb]
\centering
\subfloat[]{
\label{fig:concept1a}
\begin{tikzpicture}[yscale=.7]
\node [anchor=center,inner sep=0] at (0,0) {
\includegraphics[trim=0mm 0mm 0mm 0mm,clip=true,width=0.41\textwidth]{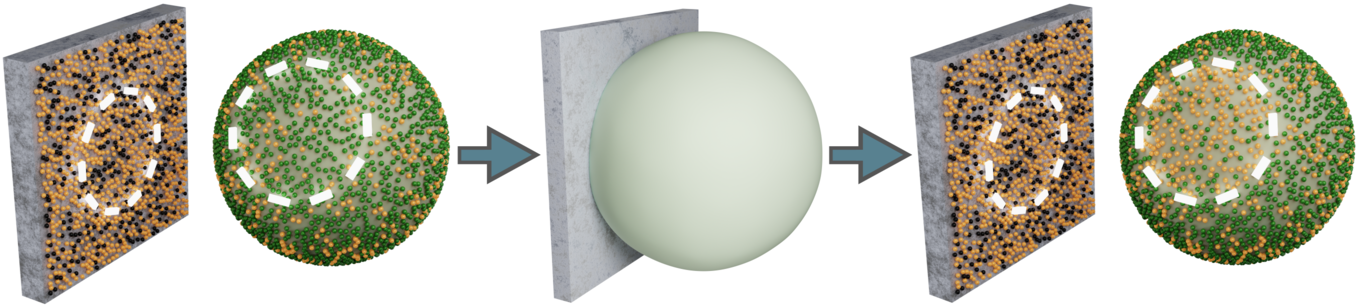}};
\draw [->,>=latex,very thick] (-.4,1.4) to [out=45,in=135] node [above,midway] {$Q_w~(\mu_w)$} (.4,1.3);
\draw [->,>=latex,very thick] (.4,-1.4) to [out=225,in=-45] node [below,midway] {$Q_i~(\mu_i,\sigma_i,\gamma_i)$} (-.4,-1.3);
\node at (.2,0) [font=\large] {$i$};
\end{tikzpicture}
}\hfill
\subfloat[]{
\label{fig:concept1b}
\begin{tikzpicture}[yscale=.7]
\node [anchor=center,inner sep=0] at (0,0) {
\includegraphics[trim=0mm 0mm 0mm 0mm,clip=true,width=0.50\textwidth]{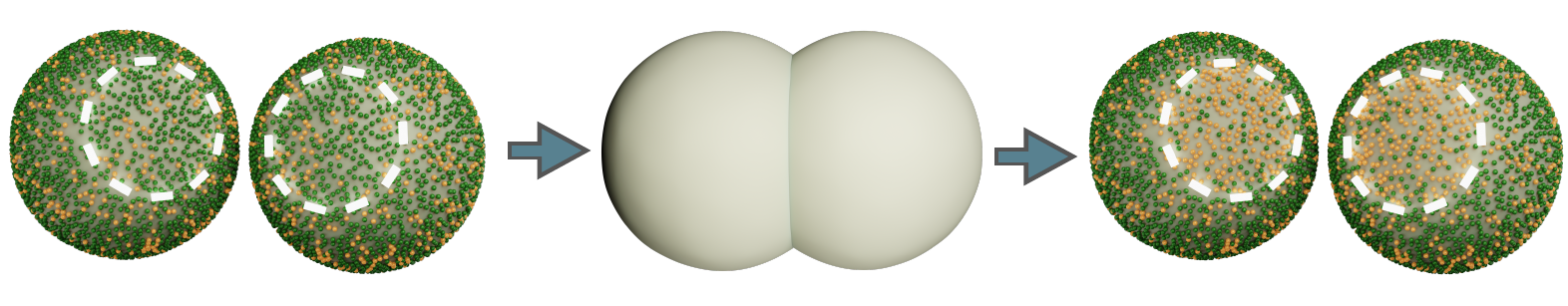}};
\draw [->,>=latex,very thick] (-.4,1.4) to [out=45,in=135] node [above,midway] {$Q_j~(\mu_j,\sigma_j,\gamma_j)$} (.4,1.3);
\draw [->,>=latex,very thick] (.4,-1.4) to [out=225,in=-45] node [below,midway] {$Q_i~(\mu_i,\sigma_i,\gamma_i)$} (-.4,-1.3);
\node at (.5,0) [font=\large] {$i$};
\node at (-.4,0) [font=\large] {$j$};
\end{tikzpicture}
}
\caption[]{%
(a) Particle and wall before (left), during (middle), and after contact (right).
The white circles indicate the contact areas.
Small dots are charge carriers on the surfaces' charging sites; their surface density is $c$.
A ratio of $\alpha$ charging sites are active.
Brown charging sites are inactive, and their charge carriers remain during contact on the surfaces.
Black sites transfer charge carriers from the wall to the particle and are thus no longer available for subsequent transfer.
As the wall is grounded and conductive, these sites instantly regenerate.
On the particle, active charging sites (green) transfer charge carriers to the wall with a probability of $p$, reducing the particle’s reservoir of transferable charge carriers.
(b) The same mechanism applied to particle-particle contact.
During impact, green charge carriers transfer between the two insulating particles, depleting the number of transferable carriers on both surfaces.}
\label{fig:concept1}
\end{figure*}
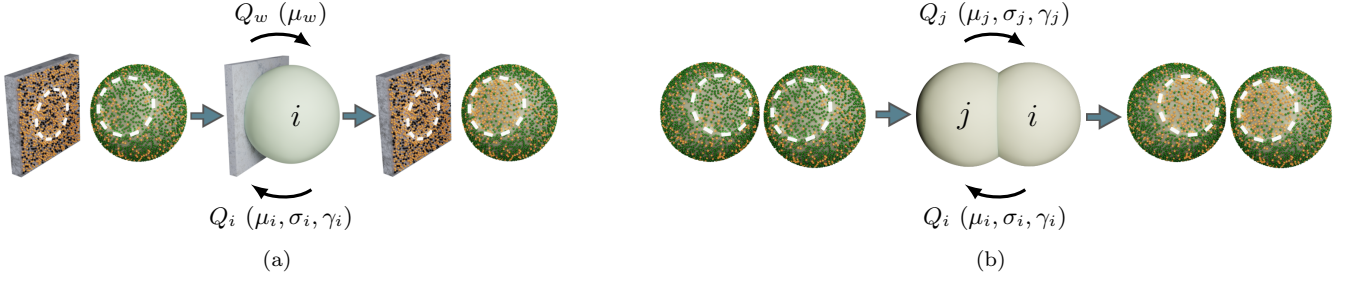

The particle charging model proposed here aims to predict the electrostatic charging of insulating particles as they contact either one another or conductive, grounded boundary walls.
\Cref{fig:concept1} illustrates the underlying concept for both interaction types.

As shown in the schematic, we denote the charge transferred from the wall to particle~$i$ as $Q_{w\rightarrow i}=Q_w$, and the charge transferred from particle~$i$ to another contacting surface, either the wall or another particle~$j$, as $Q_{i\rightarrow}=Q_i$.
Based on this notation, the impact charge, i.e., the net charge change of particle~$i$ during wall contact (\cref{fig:concept1a}) is 
\begin{subequations}
\begin{equation} 
\label{eq:qiw} 
\Delta Q_{iw} = Q_w - Q_i \, , 
\end{equation} 
and during particle-particle contact (\cref{fig:concept1b}), it is
\begin{equation} \label{eq:qij} \Delta Q_{ij} = Q_j - Q_i \, .
\end{equation}
\end{subequations}

According to the mosaic and surface state models, the contact area between a particle and another surface comprises charge-exchange sites that can be active, which means their charge carriers can transfer.
In the general formulation of the SSM, the concepts of \emph{active charging sites} and \emph{transferable charge carriers} are equivalent, and we use the terms in this paper synonymously.
Further, the SSM makes no assumptions about these sites' spatial geometry or nanostructure.

The surface density of charging sites is $c$, and a ratio of $\alpha$ of them is active.
In \cref{fig:concept1}, black markers represent the active charging sites on the wall surface, green markers the active charging sites on the particle surfaces, and brown markers denote inactive charging sites.

Measurements of charge exchange between conductive particles and surfaces~\citep{Mat03,Mat06c} show that the variation of their impact charge is minuscule compared to the variation of the impact charge between insulators.
Thus, we assume that charge transfer from the wall to the particle, $Q_w$, is deterministic.
Each of the $N_w$ active charging sites on the wall contributes a charge $\epsilon_w$ to the particle, driven by a contact potential difference or any other physical mechanism.
The model does not rely on a specific mechanism and thus remains broadly applicable.
The total transferred charge from the wall is therefore 
\begin{subequations}
\begin{equation} 
\label{eq:qw} 
Q_w = \epsilon_w N_w \sim F(\mu_w) \, , 
\end{equation} 
where $\mu_w$ is the expected value of $Q_w$.
Because the wall is conductive and grounded, the charging sites refresh immediately after contact, maintaining a constant population of active sites before and after impact, as depicted in \cref{fig:concept1a}.

In contrast, charge transfer from an insulating particle surface is assumed to be stochastic, in accordance with the mosaic model~\citep{Grosj23}.
The contact area of particle~$i$ includes $N_i$ active charging sites, each of which, with a probability of $p$, transfers a charge $\epsilon_p$ to the opposing surface.
This stochastic process follows a binomial distribution, which we approximate by a skewed normal distribution characterized by a mean $\mu_i$, standard deviation $\sigma_i$, and skewness $\gamma_i$, 
\begin{equation} 
\label{eq:qi} 
Q_i = \epsilon_p \sum_{n=1}^{N_i} \theta_n \sim G(\mu_i,\sigma_i,\gamma_i) \, ,
\end{equation} 
\end{subequations}
where $\theta_n$ is a Bernoulli random variable with two possible outcomes:
$1$ with a probability of $p$ or $0$ with a probability of $1-p$.
Unlike the conductive wall, the insulating particle's active charging sites deplete during contact, reducing the number of active charging sites after impact, as shown by the reduced green markers in \cref{fig:concept1a,fig:concept1b}.

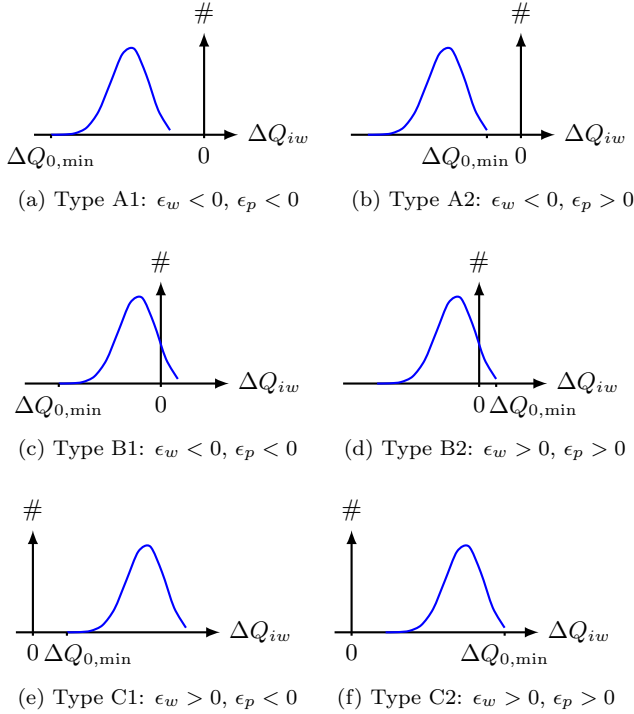
\begin{figure}[tb]
\centering
\subfloat[Type A1: $\epsilon_w<0$, $\epsilon_p<0$]{
\begin{tikzpicture}[scale=.45,font=\small]
\draw [->,>=latex,thick] (4.0,-.1) node [below] {$0$} -- (4.0,3) node [above] {\#};
\draw [->,>=latex,thick] (-1.0,0) -- (5,0) node [right] {$\Delta Q_{iw}$};
\draw [blue,thick,yscale=3.5,shift={(-.5,0)}] plot[smooth] file {fig/data/binominal-tikz.dat};
\draw [thick] (-.5,0) -- (-.5,-.1) node [below] {$\Delta Q_{0,\mathrm{min}}$};
\end{tikzpicture}
}~
\subfloat[Type A2: $\epsilon_w<0$, $\epsilon_p>0$]{
\begin{tikzpicture}[scale=.45,font=\small]
\draw [->,>=latex,thick] (4.0,-.1) node [below] {$0$} -- (4.0,3) node [above] {\#};
\draw [->,>=latex,thick] (-1.0,0) -- (5,0) node [right] {$\Delta Q_{iw}$};
\draw [blue,thick,yscale=3.5,shift={(-.5,0)}] plot[smooth] file {fig/data/binominal-tikz.dat};
\draw [thick] (3.0,0) -- (3.0,-.1) node [below,xshift=-8] {$\Delta Q_{0,\mathrm{min}}$};
\end{tikzpicture}
}\\
\subfloat[Type B1: $\epsilon_w<0$, $\epsilon_p<0$]{
\begin{tikzpicture}[scale=.45,font=\small]
\draw [->,>=latex,thick] (2.5,-.1) node [below] {$0$} -- (2.5,3) node [above] {\#};
\draw [->,>=latex,thick] (-1.5,0) -- (4.5,0) node [right] {$\Delta Q_{iw}$};
\draw [blue,thick,yscale=3.5,shift={(-.5,0)}] plot[smooth] file {fig/data/binominal-tikz.dat};
\draw [thick] (-.5,0) -- (-.5,-.1) node [below] {$\Delta Q_{0,\mathrm{min}}$};
\end{tikzpicture}
}~
\subfloat[Type B2: $\epsilon_w>0$, $\epsilon_p>0$]{
\begin{tikzpicture}[scale=.45,font=\small]
\draw [->,>=latex,thick] (2.5,-.1) node [below] {$0$} -- (2.5,3) node [above] {\#};
\draw [->,>=latex,thick] (-1.5,0) -- (4.5,0) node [right] {$\Delta Q_{iw}$};
\draw [blue,thick,yscale=3.5,shift={(-.5,0)}] plot[smooth] file {fig/data/binominal-tikz.dat};
\draw [thick] (3.0,0) -- (3.0,-.1) node [below,xshift=14] {$\Delta Q_{0,\mathrm{min}}$};
\end{tikzpicture}
}\\
\subfloat[Type C1: $\epsilon_w>0$, $\epsilon_p<0$]{
\begin{tikzpicture}[scale=.45,font=\small]
\draw [->,>=latex,thick] (-1.5,-.1) node [below] {$0$} -- (-1.5,3) node [above] {\#};
\draw [->,>=latex,thick] (-2,0) -- (4.0,0) node [right] {$\Delta Q_{iw}$};
\draw [blue,thick,yscale=3.5,shift={(-.5,0)}] plot[smooth] file {fig/data/binominal-tikz.dat};
\draw [thick] (-.5,0) -- (-.5,-.1) node [below,xshift=8] {$\Delta Q_{0,\mathrm{min}}$};
\end{tikzpicture}
}~
\subfloat[Type C2: $\epsilon_w>0$, $\epsilon_p>0$]{
\begin{tikzpicture}[scale=.45,font=\small]
\draw [->,>=latex,thick] (-1.5,-.1) node [below] {$0$} -- (-1.5,3) node [above] {\#};
\draw [->,>=latex,thick] (-2,0) -- (4.0,0) node [right] {$\Delta Q_{iw}$};
\draw [blue,thick,yscale=3.5,shift={(-.5,0)}] plot[smooth] file {fig/data/binominal-tikz.dat};
\draw [thick] (3.0,0) -- (3.0,-.1) node [below] {$\Delta Q_{0,\mathrm{min}}$};
\end{tikzpicture}
}
\caption[]{%
Three possible impact charge distributions (Types A, B, and C) for contacts between fresh, uncharged, insulating particles and a conductive wall (sketched for negative skewness). 
Each distribution type allows two distinct interpretations (labeled 1 and 2) regarding the polarity of the transferred charge carriers $\epsilon_w$ from the wall and $\epsilon_p$ from the particle.
Types~A1 and B1 appear to be most representative of the materials in this study; 
nevertheless, the model is constructed in a general form.}
\label{fig:concept2}
\end{figure}

We assume that $\epsilon_w$ and $\epsilon_p$ are constant for a given powder, an assumption that aligns with \citet{Grosj23}. 
Beyond this, the model does not require specification of the charge-exchange sites' size, shape, or nanoscopic structure. 
The parameters $\epsilon_w$ and $\epsilon_p$ represent quantized transferable charge carriers and are integer multiples of the elementary charge. 

The polarities of $\epsilon_w$ and $\epsilon_p$ determine six distinct net charge distributions after impact (see \cref{fig:concept2}). 
When $\epsilon_w$ and $\epsilon_p$ have opposite signs, the resulting distributions are strictly unipolar: 
either always negative (Type~A2) or always positive (Type~C1). 
If $\epsilon_w$ and $\epsilon_p$ have the same sign, the net impact charge can still be unipolar, provided the wall’s contribution dominates (Types~A1 and~C2).
However, if the particle’s contribution is comparable, the distributions span both polarities, resulting in bipolar distributions (Types~B1 and~B2).

The charge histograms can identify the principal distribution types (A, B, or C). 
Distinguishing subtypes (1 or 2) requires further analysis. 
In our particle impact experiments, the charge distribution's tails extend into the negative domain. 
Given that electrons are the most probable charge carriers from grounded conductors suggests $\epsilon_w < 0$. 
If the distribution spans both signs, then $\epsilon_p < 0$ as well (i.e., Type~B1). 
If the distribution is unipolar negative, the charge carriers are probably the same polarity but the wall contribution dominates;
thus, Type~A1 is most likely, but A2 cannot be entirely excluded.
Therefore, Types~A1 and~B1 appear to be most representative of the materials examined in this study.
Importantly, the model remains general in form and does not depend on the identity of the specific charge carriers.

\subsection{Scaling the statistical parameters: $\mu_w$, $\mu_i$, $\sigma_i$, $\gamma_i$}
\label{sec:scaling}

\begin{figure*}[tb]
\centering
\subfloat[]{
\begin{tikzpicture}[scale=.5]
\draw [->,>=latex,thick] (0,-.1) node [below] {$0$} -- (0,3.3) node [above] {\#};
\draw [->,>=latex,thick] (-3.5,0) -- (4.5,0) node [below] {$\Delta Q_{iw}$};
\draw [blue,yscale=3.5,shift={(0,0)}] plot[smooth] file {fig/data/binominal-tikz.dat};
\node at (3.9,1.1) [blue] {$-Q_i$};
\draw [blue,dotted,thin,xscale=-1,yscale=3.5,shift={(0,0)}] plot[smooth] file {fig/data/binominal-tikz.dat};
\node at (-3.9,1.1) [blue] {$Q_i$};
\draw [thick,yscale=3.5,shift={(-3.0,0)}] plot[smooth] file {fig/data/binominal-tikz.dat};
\draw [blue] (-3.0,0) -- (-3.0,2.5);
\draw [blue,dotted] (-3.0,2.5) -- (-3.0,2.8) node [above] {\tikzmark{a1}$\mu_{w0}$};
\draw [thin,dashed] (-.7,3) -- (-.7,-.1) node [below] {$\mu_0$};
\draw [thin,dashed] (-1.3,3) -- (-1.3,1);
\draw [thin] (-.7,2.9) to node [above,midway] {$\sigma_0$} (-1.3,2.9);
\node at (-3,0) [below,xshift=-4] {$\Delta Q_{0,\mathrm{min}}$};
\draw [thin,dashed,blue] (2.3,3.0) -- (2.3,-.1) node [below] {\tikzmark{a2}$-\mu_{i0}$};
\draw [thin,dashed,blue] (2.9,3.0) -- (2.9,1.0);
\draw [thin,blue] (2.3,2.9) to node [above,midway] {\tikzmark{a3}$\sigma_0$} (2.9,2.9);
\end{tikzpicture}
}\quad
\subfloat[]{
\begin{tikzpicture}[scale=.5]
\draw [->,>=latex,thick] (0,-.1) node [below] {$0$} -- (0,3.3) node [above] {\#};
\draw [->,>=latex,thick] (-3.5,0) -- (4.5,0) node [below] {$\Delta Q_{iw}$};
\draw [blue,yscale=3.5,shift={(0,0)}] plot[smooth] file {fig/data/binominal-tikz.dat};
\node at (3.9,1.1) [blue] {$-Q_i$};
\draw [thick,yscale=3.5,shift={(-3.0,0)}] plot[smooth] file {fig/data/binominal-tikz.dat};
\draw [blue] (-3.0,0) -- (-3.0,2.5);
\draw [blue,dotted] (-3.0,2.5) -- (-3.0,2.8) node [above] {\tikzmark{b1}$\mu_w$};
\draw [thin,dashed,blue] (2.3,3.0) -- (2.3,-.1) node [below] {\tikzmark{b2}$-\mu_i$};
\draw [thin,dashed,blue] (2.9,3.0) -- (2.9,1.0);
\draw [thin,blue] (2.3,2.9) to node [above,midway] {\tikzmark{b3}$\sigma_i$} (2.9,2.9);
\end{tikzpicture}
}\quad
\subfloat[]{
\begin{tikzpicture}[scale=.5]
\draw [->,>=latex,thick] (0,-.1) node [below] {$0$} -- (0,3.3) node [above] {\#};
\draw [->,>=latex,thick] (-4.5,0) -- (4.5,0) node [below] {$\Delta Q_{ij}$};
\draw [blue,yscale=3.5,shift={(0,0)}] plot[smooth] file {fig/data/binominal-tikz.dat};
\node at (3.9,1.1) [blue] {$-Q_i$};
\draw [blue,xscale=-1,yscale=3.5,shift={(0,0)}] plot[smooth] file {fig/data/binominal-tikz.dat};
\node at (-3.9,1.1) [blue] {$Q_j$};
\def\mua{0};
\def\sigmaa{.85};
\draw [domain=-3.5:3.5, smooth, thick, variable=\x] plot ({\x}, {4/(\sigmaa*sqrt(2*pi))*exp(-((\x-\mua)^2)/(2*\sigmaa^2))});
\draw [thin,dashed,blue] (2.3,3.0) -- (2.3,-.1) node [below] {\tikzmark{c1}$-\mu_i$};
\draw [thin,dashed,blue] (2.9,3.0) -- (2.9,1.0);
\draw [thin,blue] (2.3,2.9) to node [above,midway] {\tikzmark{c2}$\sigma_i$} (2.9,2.9);
\draw [thin,dashed,blue] (-2.3,3.0) -- (-2.3,-.1) node [below] {\tikzmark{c3}$\mu_j$};
\draw [thin,dashed,blue] (-2.9,3.0) -- (-2.9,1.0);
\draw [thin,blue] (-2.3,2.9) to node [above,midway] {\tikzmark{c4}$\sigma_j$} (-2.9,2.9);
\end{tikzpicture}
\begin{tikzpicture}[overlay, remember picture,shorten <=1mm,->,>=latex,Green,opacity=.7]
\draw ([xshift=1.5ex,yshift=1.5ex] pic cs:a1) to [out=15,in=165] ([xshift=1.5ex,yshift=1.5ex] pic cs:b1);
\draw ([xshift=1.5ex,yshift=1.5ex] pic cs:a3) to [out=15,in=165] ([xshift=1.5ex,yshift=1.5ex] pic cs:b3);
\draw ([xshift=1.5ex,yshift=1.5ex] pic cs:a3) to [out=15,in=165] ([xshift=1.5ex,yshift=1.5ex] pic cs:c2);
\draw ([xshift=1.5ex,yshift=1.5ex] pic cs:a3) to [out=15,in=165] ([xshift=1.5ex,yshift=1.5ex] pic cs:c4);
\draw ([xshift=1.5ex,yshift=-1.0ex] pic cs:a2) to [out=-15,in=-165] ([xshift=1.5ex,yshift=-1.0ex] pic cs:b2);
\draw ([xshift=1.5ex,yshift=-1.0ex] pic cs:a2) to [out=-15,in=-165] ([xshift=1.5ex,yshift=-1.0ex] pic cs:c1);
\draw ([xshift=1.5ex,yshift=-1.0ex] pic cs:a2) to [out=-15,in=-165] ([xshift=1.5ex,yshift=-1.0ex] pic cs:c3);
\end{tikzpicture}
}
\caption[]{%
(a) The impact charge distribution from reference experiments (black, Type~B1) provides the statistical parameters $\mu_0$, $\sigma_0$, $\gamma_0$, and $\Delta Q_{0,\mathrm{min}}$.
We decompose this distribution into the transfers (blue) from the wall and the particle.
The SSM scales these parameters (green arrows) to each contact:
(b) For particle-wall contacts, to $\mu_w$, $\mu_i$, $\sigma_i$, and $\gamma_i$;
(c) For particle-particle contacts, to $\mu_i$, $\sigma_i$, $\gamma_i$, $\mu_j$, $\sigma_j$, and $\gamma_j$.
(For clarity, skewness values $\gamma_0$, $\gamma_i$, and $\gamma_j$ are not shown in the plots.)
Superimposing the contributions gives the impact charge distribution (black) per contact.}
\label{fig:variables}
\end{figure*}
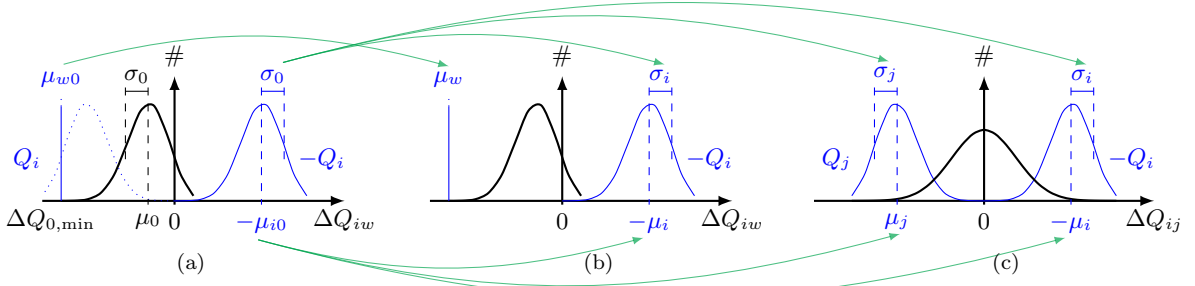

As described in \cref{sec:reference-impact}, particle impact experiments provide, for a defined reference condition (denoted by the subscript 0), the statistical parameters of the impact charge distribution.
These parameters, shown in \cref{fig:variables}, serve as the basis for scaling the charge transfer distributions to varying impact conditions.

According to \cref{eq:qw}, a grounded conductive wall transfers the charge
\begin{subequations}
\begin{equation}
\label{eq:muw}
\mu_w = \epsilon_w N_w 
\end{equation}
to a particle.
The charge transfer from an insulating particle follows, based on \cref{eq:qi}, a binomial distribution, characterized by the mean
\begin{equation}
\label{eq:mui}
\mu_i = \epsilon_p N_i p \, , 
\end{equation} 
standard deviation
\begin{equation}
\label{eq:sigmai}
\sigma_i = \epsilon_p \sqrt{N_i p(1-p)} \, ,
\end{equation}
and skewness
\begin{equation}
\label{eq:gammai}
\gamma_i = \dfrac{1-2p}{\sqrt{N_i p (1-p)}} \, ,
\end{equation}
\end{subequations}
where $p$ is the probability that an active site transfers charge.
\Cref{eq:mui,eq:sigmai,eq:gammai} hold analogously for particle $j$ in a particle-particle contact.

The problem in solving \cref{eq:muw,eq:mui,eq:sigmai,eq:gammai} lies in the nanoscopic parameters $\epsilon_w$, $\epsilon_p$, $p$, $N_w$, and $N_i$, which are not directly measurable.
The SSM circumvents this challenge by scaling the parameters of the reference impact. 
For the reference impact, the same expressions become
\begin{subequations}
\begin{equation}
\label{eq:muw0}
\mu_{w0} = \epsilon_w N_{w0} \, ,
\end{equation}
\begin{equation}
\label{eq:mui0}
\mu_{i0} = \epsilon_p N_0 p \, , 
\end{equation}
\begin{equation}
\label{eq:sigma0}
\sigma_{i0} = \epsilon_p \sqrt{N_0 p(1-p)} = \sigma_0 \, ,
\end{equation}
\begin{equation}
\label{eq:gamma0}
\gamma_{i0} =   \dfrac{1-2p}{\sqrt{N_0 p (1-p)}} = -\gamma_0 \, ,
\end{equation}
 \end{subequations}
where $N_{w0}$ and $N_0$ represent the active charging sites for the wall and the particle, respectively, during the reference impact.
Because all variation of the impact charge stems from the insulating particle, the standard deviation of the charge transfer from the particle equals the overall standard deviation of the impact charge (\cref{eq:sigma0}).
The skewness changes sign (\cref{eq:gamma0}) because of the sign convention in \cref{eq:qiw}.

From these definitions, we derive the scaling laws for the particle-wall and particle-particle charge transfer parameters.
Taking the ratio of \cref{eq:muw} to \cref{eq:muw0} gives the scaled wall-to-particle mean charge transfer, 
\begin{equation}
\label{eq:muw-scale}
\mu_w= \mu_{w0} \dfrac{N_w}{N_{w0}} \, .
\end{equation}
Likewise, the scaled mean of the particle-to-surface transfer follows from the ratio of \cref{eq:mui} to \cref{eq:mui0},
\begin{equation}
\label{eq:mui-scale}
\mu_i= \mu_{i0} \dfrac{N_i}{N_0} \, .
\end{equation}
Besides the ratios of the charging sites ($N_i/N_0$ and $N_w/N_{w0}$), \cref{eq:muw-scale,eq:mui-scale} contain the unknowns $\mu_{w0}$ and $\mu_{i0}$.

To resolve $\mu_{i0}$, we use the mean of the charge balance \cref{eq:qiw} for the reference impact, yielding 
\begin{equation}
\label{eq:mui0}
\mu_{i0} = \mu_{w0} - \mu_0 \, .
\end{equation}
This leaves the remaining unknown $\mu_{w0}$, the wall-to-particle charge transfer in the reference condition.

We approximate $\mu_{w0}$ by the minimum observed impact charge in the reference experiment, $\Delta Q_{0,\mathrm{min}}$.
This estimate assumes that the smallest recorded impact charge corresponds to zero particle-to-wall charge transfer ($Q_i=0$).
This approximation is asymptotically valid in the limit of a large experimental sample size,
\begin{equation}
\label{eq:approx-muw0}
\mu_{w0} = \lim_{M \to \infty} \left( \Delta Q_{0,\mathrm{min}} \right) \approx \Delta Q_{0,\mathrm{min}} \, .
\end{equation}
Since our experimental dataset may not fully reach this limit, we test the sensitivity of our simulations to this assumption in \cref{sec:err}.

Finally, using approximation~\ref{eq:approx-muw0} with \cref{eq:muw-scale} yields the scaled charge transfer from the wall to a particle as 
\begin{subequations}
\begin{equation}
\label{eq:muw-scale2}
\mu_w = \Delta Q_{0,\mathrm{min}} \dfrac{N_w}{N_{w0}}
\end{equation}
and with \cref{eq:mui-scale,eq:mui0} the scaled mean charge transfer from a particle to another surface as
\begin{equation}
\label{eq:mui-scale2}
\mu_i= \left( \Delta Q_{0,\mathrm{min}} - \mu_0 \right) \dfrac{N_i}{N_0} \, .
\end{equation}
Further, the scaled standard deviation is obtained from \cref{eq:sigmai} and \cref{eq:sigma0},
\begin{equation}
\label{eq:sigmai-scale}
\sigma_i= \sigma_0 \sqrt{\dfrac{N_i}{N_0}} \, ,
\end{equation}
and the scaled skewness from \cref{eq:gammai} and \cref{eq:gamma0} is
\begin{equation} 
\label{eq:gammai-scale} 
\gamma_i= -\gamma_0 \sqrt{\dfrac{N_0}{N_i}} \, . 
\end{equation}
\end{subequations}

\Cref{eq:gammai-scale,eq:sigmai-scale,eq:mui-scale2,eq:muw-scale2} form the scaling laws to define the statistical properties of particle-wall charge transfer.
For particle-particle charge transfer, the corresponding scaling laws follow \cref{eq:gammai-scale,eq:sigmai-scale,eq:mui-scale2} for particle $i$ and the analogous expressions for particle $j$.
The ratios $N_i/N_0$ and $N_w/N_{w0}$ scale the contact-specific statistics from the reference condition and are further addressed in the next section.

Using \cref{eq:gammai-scale,eq:sigmai-scale,eq:mui-scale2} for particle-particle charge transfer implies that the charge-transfer distribution for same-material contacts can be inferred from the distribution measured for different-material contacts.
We test this assumption in \cref{sec:test-scaling}.

The wall's charge transfer, $Q_w$, remains deterministic via \cref{eq:qw}, while the particle’s contribution, $Q_i$, follows a binomial distribution.
For sufficiently large $N_i$, a binomial distribution is well-approximated by the skewed normal distribution.
Thus, $Q_i$ is sampled from a skewed normal distribution with the probability density function 
\begin{equation}
\label{eq:skew-normal}
\begin{split}
 g&(Q_i;\xi_i,\omega_i,\delta_i) = \dfrac{1}{\omega_i \sqrt{2 \pi}} ~ \mathrm{e}^{-(Q_i-\xi_i)^2/(2\omega_i^2)} \\
 &\cdot \left[1+\mathrm{erf}\left(\dfrac{\delta_i(Q_i-\xi_i)}{\sqrt{2}\omega_i}\right) \right]
~\text{for}~~ 0 \le Q_i \le Q_{i,\mathrm{max}} \, , 
\end{split} 
\end{equation}
where the shape parameters, $\xi_i$, $\omega_i$, and $\delta_i$, are functions of the distribution's statistical parameters $\mu_i$, $\sigma_i$, and $\gamma_i$.

Since the skewed normal distribution is continuous, its tails extend to $\pm \infty$, while $Q_i$ is physically limited by the finite number of active charging sites.
The lower bound $Q_i = 0$ corresponds to no charge transfer; the upper bound $Q_{i,\mathrm{max}}$ corresponds to all active sites transferring charge.
As shown in Appendix~\ref{sec:qimax}, this upper limit is given by the closed-form expression 
\begin{equation}
\label{eq:qimax}
Q_{i,\mathrm{max}} = \dfrac{\mu_i \left( \gamma_i \mu_i - 2 \sigma_i \right)}{\gamma_i \mu_i - \sigma_i} \, .
\end{equation} 
To ensure the total probability of the truncated distribution remains unity, we normalize \cref{eq:skew-normal}.
To generate stochastic realizations of $Q_i$, we apply an inversion sampling method using uniformly distributed pseudorandom numbers.

\subsection{Active charging sites: $N_i/N_0$ and $N_w/N_{w0}$}
\label{sec:N}

The scaling relations \labelcref{eq:gammai-scale,eq:sigmai-scale,eq:mui-scale2,eq:muw-scale2} derived in the previous section require approximating the ratios $N_i/N_0$ and $N_w/N_{w0}$.
A key advantage of the SSM is that it relies only on these relative quantities, rather than requiring knowledge of the absolute number of charging sites.

If the particle and wall surfaces are homogeneous, the product of the contact area, $A$, and the surface density of active charging sites gives the number of active charging sites. 
At the wall, the charging sites regenerate instantaneously, so their surface density remains constant.
In contrast, the initial surface density on the particle, $c_0$, depletes over time due to charge transfer, so that $0 \le c/c_0 \le 1$.

Additionally, the local electric field around the particle can act as a force on the charge carriers against their path from one surface to the other.
Then, the local electric field reduces the ratio $\alpha$ of active charging sites, such that $0 \le \alpha/\alpha_0 \le 1$, where $\alpha$ is $\alpha_w$ at the wall and $\alpha_i$ on the particle surface and $\alpha_0$ is the ratio in the absence of a surrounding electric field.

Combining these effects, the number of active charging sites relative to the reference condition scales as
\begin{equation}
\label{eq:N}
\dfrac{N_w}{N_{w0}} = \dfrac{A}{A_0} \dfrac{\alpha_w}{\alpha_0}
\quad\text{and}\quad
\dfrac{N_i}{N_0} = \dfrac{A}{A_0} \dfrac{\alpha_i}{\alpha_0} \dfrac{c}{c_0} \, .
\end{equation}
The next sections develop physical approximations for the ratios $A/A_0$ (\cref{sec:A}), $\alpha_w/\alpha_0$ and $\alpha_i/\alpha_0$ (\cref{sec:alpha}), and $c/c_0$ (\cref{sec:c}).

\subsection{Contact surface area: $A/A_0$}
\label{sec:A}

We model the ratio $A/A_0$ using Hertzian elastic contact theory and consider additionally the effect of surface roughness and the contact mode.
For particle-wall interactions, i.e., a smooth spherical particle impacting a flat surface with the normal velocity $v_{\mathrm{n},i}$, the contact area is~\citep{John80}
\begin{equation}
\label{eq:aw-hertz}
A=A_w= \pi r_i^2 \left( \dfrac{5}{8} \pi \rho_\mathrm{p} (1+k_e) v_{\mathrm{n},i}^2 (\chi_p + \chi_w )\right)^{2/5} \, ,
\end{equation}
where $k_e$ is the particle's restitution coefficient.
The elasticity parameters of the wall, $\chi_w$, and particle, $\chi_p$, are given by $(1-a^2)/b$, where $a$ and $b$ are their Poisson ratios and Young moduli, and usually $\chi_p \gg \chi_w$.

The ratio of the contact area to that of the reference impact becomes
\begin{equation}
\label{eq:aw}
\dfrac{A}{A_0} = 
\dfrac{A_w}{A_0} = 
\beta_\mathrm{c}
\left(\dfrac{r_i}{r_0}\right)^{2\beta_\mathrm{r}}
\left(\dfrac{v_{\mathrm{n},i}}{v_0}\right)^{4/5} \, ,
\end{equation}
where $r_0$ and $v_0$ are the reference particle radius and normal impact velocity.
The coefficient $\beta_\mathrm{c}$ accounts for the effective increase in contact area due to rolling and sliding motion during wall contacts.
For flows parallel to walls, such as in pneumatic transport systems, the tangential impact velocity can exceed the normal component by up to three orders of magnitude.
Under such conditions, we assume that the particle rolls once over its circumference during contact and further increases the effective contact area due to sliding, leading to $\beta_\mathrm{c} = 4 \beta_a \, r_i (\pi / A_w)^{1/2}$ with $\beta_a=1$.
Only for simulations 1 and 2 (cf.~\cref{tab:cfd}), $\beta_a$ is larger than unity to shorten the computation duration.

The coefficient $\beta_\mathrm{r}$ accounts for the effect of surface roughness on the contact area.
Note that $\beta_\mathrm{r}$ compensates for surface roughness in the ratio $A_w/A_0$, which is a minor influence compared to the absolute influence of roughness on $A_w$.
Based on our earlier findings~\citep{Jan24a}, we adopt $\beta_\mathrm{r} = 2.4$.

For particle-particle collisions, Hertz's theory provides the expression~\citep{Soo71}
\begin{equation}
\label{eq:aij-hertz}
A=A_{ij}= \dfrac{\pi \, r_i^2 \, r_j^2}{r_i + r_j}  \left( \dfrac{5}{8} \pi \rho_\mathrm{p} (1+k_e) v_{ij}^2 \dfrac{\sqrt{r_i + r_j}}{r_i^3 + r_j^3} \chi_p \right)^{2/5}
\end{equation}
for the contact area, where $v_{ij}$ denotes the relative impact velocity of the colliding particles with radii $r_i$ and $r_j$.
Accordingly, the ratio to the reference contact area is
\begin{equation}
\dfrac{A}{A_0} = 
\dfrac{A_{ij}}{A_0} = 
\left(
\dfrac{r_i^5 \, r_j^5}{r_0^5 \left(r_i+r_j\right)^2 (r_i^3+r_j^3)}
\dfrac{v^2_{ij}}{v^2_0}
\dfrac{\chi_\mathrm{p}}{\chi_\mathrm{p}+\chi_\mathrm{w}} 
\right)^{2\beta_\mathrm{r}/5} \, .
\end{equation}
Since $\chi_p \gg \chi_w$, the above simplifies to
\begin{equation}
\label{eq:aij}
\dfrac{A}{A_0} = 
\dfrac{A_{ij}}{A_0} = 
\left(
\dfrac{r_i^5 \, r_j^5}{r_0^5 \left(r_i+r_j\right)^2 (r_i^3+r_j^3)}
\dfrac{v^2_{ij}}{v^2_0}
\right)^{2\beta_\mathrm{r}/5} \, .
\end{equation}

In summary, \cref{eq:aw,eq:aij} provide the scaling laws for the contact area used in \cref{eq:N}.
The present model is based on classical Hertzian theory and additionally accounts for surface roughness and contact mode.
It can be systematically extended to include more complex contact mechanics, for example, of non-spherical particles or varying contact modes~\citep{Ire08,Ire12}, all of which can be incorporated into the scaling ratio $A/A_0$.

Overall, comparing \cref{eq:aw} and \cref{eq:aij} to \cref{eq:aw-hertz} and \cref{eq:aij-hertz} demonstrates once more the advantage of the SSM building throughout on ratios instead of absolute quantities, which cancels out most material-dependent parameters.

\subsection{Activity ratio: $\alpha/\alpha_0$}
\label{sec:alpha}

In the reference experiment, particles are initially uncharged before impact and shielded from external electric fields.
As a particle accumulates charge, the surrounding electric field ${\bm E}$ increases, hindering charge transfer in subsequent collisions.
This suppression of charge transfer is captured by the ratio of active charging sites, $\alpha$, generally expressed as~\citep{Zhang24}
\begin{equation}
\label{eq:alpha}
\alpha = \alpha_0 \left( 1 - \mathrm{sgn}(\epsilon) \dfrac{{\bm E}}{|E_\mathrm{sat}|} \cdot \dfrac{{\bm n}_\mathrm{contact}}{\|{\bm n}_\mathrm{contact}\|} \right) 
~\text{with}~ 0 \le \alpha \le \alpha_0 \\
\end{equation}
where $\alpha_0$ is the activity ratio of a particle surface without an incident electric field.
The sign function accounts for the directionality of the field with respect to the charge carrier's polarity.

Further, ${\bm E}$ is the surrounding electric field at the exact point of contact, generated by all charged particles in the system and their mirror charges at the walls, given in the global coordinate system.

The magnitude of the saturation field strength, $|E_\mathrm{sat}|$, is the absolute of the field strength at which all sites cease to transfer charge.
For a contact of a single charged particle with a conducting surface,
\begin{equation}
\label{eq:esat}
|E_\mathrm{sat}| = \dfrac{1}{2 \pi \varepsilon} \dfrac{|Q_\mathrm{sat}|}{r_0^2} \, ,
\end{equation}
depends on the corresponding magnitude of the particle saturation charge, which we approximate by 
\begin{equation}
\label{eq:qsat}
|Q_\mathrm{sat}| = C_\mathrm{sat} \, r_0^2 \, .
\end{equation}
Here, $C_\mathrm{sat} = 500 \, \upmu\mathrm{C \, m^{-2}}$ is a fitting parameter based on numerical simulations for air at atmospheric pressure and unit permittivity~\citep{Mat03}, which also aligns with experimental data~\citep{Mat18}.
Substituting \cref{eq:qsat} into \cref{eq:esat} yields
\begin{equation}
\label{eq:esat2}
|E_\mathrm{sat}| = \dfrac{C_\mathrm{sat}}{2 \pi \varepsilon} \, .
\end{equation}

The normal vector of the contact plane is ${\bm n}_\mathrm{contact}$.
For particle-wall collisions, ${\bm n}_\mathrm{contact}$ is the wall's normal vector.
For particle-particle collisions, ${\bm n}_\mathrm{contact}$ is the vector connecting the center points of the two particles.
The length of the normal vector is $\|{\bm n}_\mathrm{contact}\|$.

\begin{figure*}[tb]
\centering
\subfloat[]{
\label{fig:spca}
\includegraphics[trim=0mm 0mm 0mm 0mm,clip=true,height=.26\textwidth]{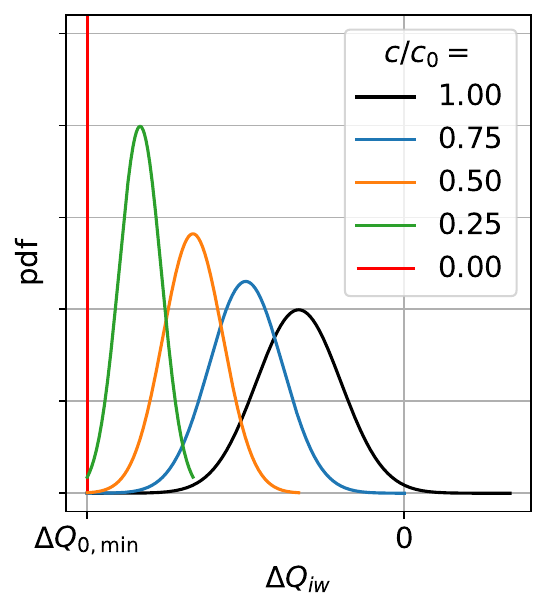}
}
\subfloat[]{
\label{fig:spcb}
\includegraphics[trim=0mm 0mm 0mm 0mm,clip=true,height=.26\textwidth]{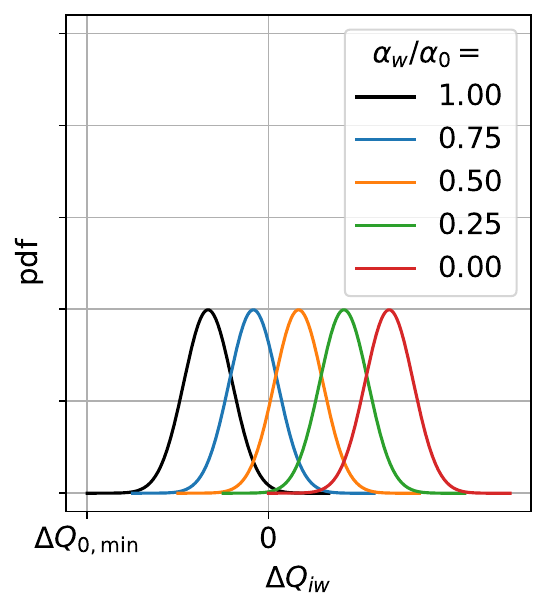}
}
\subfloat[]{
\label{fig:spcc}
\includegraphics[trim=0mm 0mm 0mm 0mm,clip=true,height=.26\textwidth]{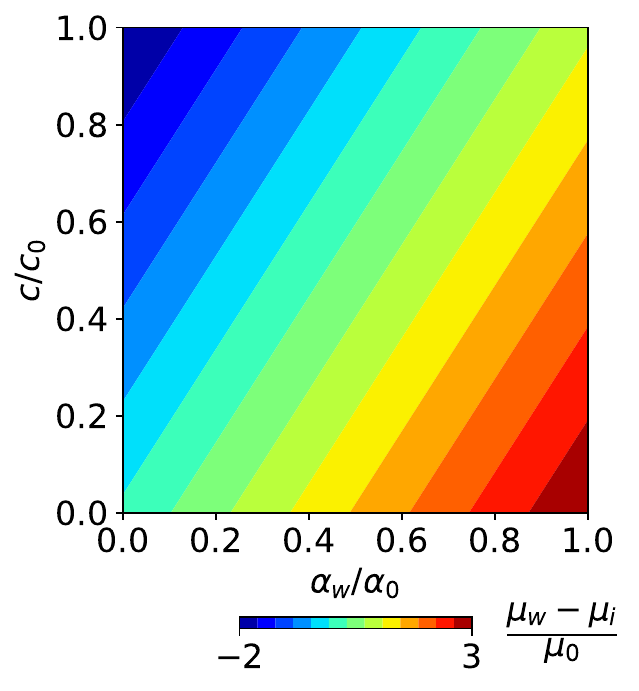}
}
\subfloat[]{
\label{fig:spcd}
\includegraphics[trim=0mm 0mm 0mm 0mm,clip=true,height=.26\textwidth]{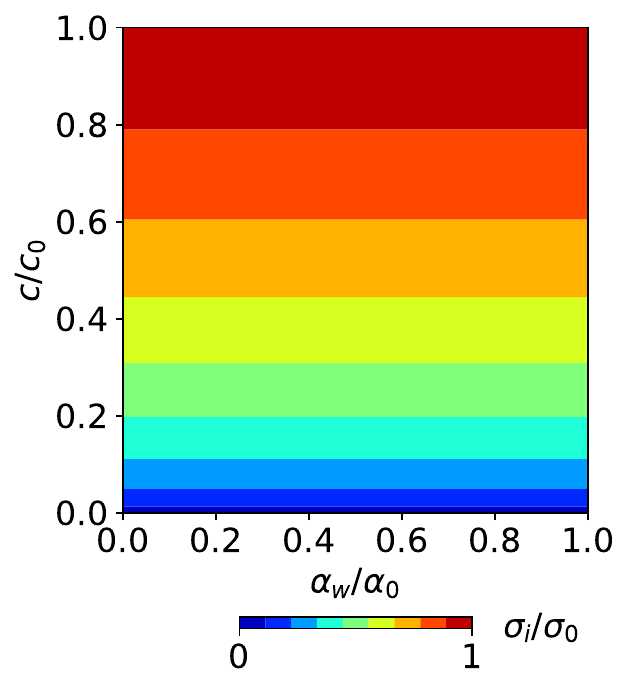}
}
\caption[]{%
Statistical parameters of repeated particle-wall impacts predicted by the SSM ($A/A_0 = \text{const}$, $\gamma_0 = 0$, Type~B1; see \cref{fig:concept2}).  
(a) Probability distributions of the impact charge for uncharged particles ($\alpha_w/\alpha_0 = 1$) with depleting charging sites on the surface.  
(b) Distributions for particles maintaining their initial charging site density ($c/c_0 = 1$) while their charge increases.  
(c) Mean and (d) standard deviation of the impact charge as functions of $\alpha_w/\alpha_0$ and $c/c_0$.}
\label{fig:spc}
\end{figure*}

The transferable charge carrier in the sign function of \cref{eq:alpha} is expressed in quantities of the reference impact;
for the charge transfer from a wall to a particle by
\begin{subequations}
\begin{equation}
\mathrm{sgn}(\epsilon)=\mathrm{sgn}(\epsilon_p)=\mathrm{sgn}(\Delta Q_{0,\mathrm{min}}-\mu_0)
\end{equation}
and from a particle to another surface by
\begin{equation}
\mathrm{sgn}(\epsilon)=\mathrm{sgn}(\epsilon_w)=\mathrm{sgn}(\Delta Q_{0,\mathrm{min}}) \, .
\end{equation}
\end{subequations}

Then, the final scaling relations for the active site ratios used in \cref{eq:N} read for a wall,
\begin{subequations}
\begin{equation}
\label{eq:alphaw}
\dfrac{\alpha}{\alpha_0} = \dfrac{\alpha_w}{\alpha_0} = 1 + \mathrm{sgn}(\Delta Q_{0,\mathrm{min}})\dfrac{{\bm E}}{|E_\mathrm{sat}|} \cdot \dfrac{{\bm n}_\mathrm{contact}}{\|{\bm n}_\mathrm{contact}\|} \, ,
\end{equation}
and for a particle contacting a wall or another particle,
\begin{equation}
\label{eq:alphai}
\dfrac{\alpha}{\alpha_0} = \dfrac{\alpha_i}{\alpha_0} = 1 - \mathrm{sgn}(\Delta Q_{0,\mathrm{min}}-\mu_0)\dfrac{{\bm E}}{|E_\mathrm{sat}|} \cdot \dfrac{{\bm n}_\mathrm{contact}}{\|{\bm n}_\mathrm{contact}\|} \, ,
\end{equation}
\end{subequations}
with $0 \le \alpha_w/\alpha_0 \le 1$ and $0 \le \alpha_i/\alpha_0 \le 1$.

\subsection{Charging site density: $c/c_0$}
\label{sec:c}

A particle undergoes $M$ contacts with other surfaces, which occur at the times of $(t_m)^M_{m=1}$.
Before the first contact ($t<t_1$), the surface density of charging sites remains unchanged from its initial value, $c_0$, which equals the density on the surface of the levitated particle in the reference experiment.
During the $m$th collision, the density reduces from $c_{m-1}$ to $c_m$ and the relative density reduces according to
\begin{equation}
\label{eq:dc}
\Delta\left(\dfrac{c}{c_0}\right)_m
= 
\left( \dfrac{c_m-c_{m-1}}{c_0} \right)
=
- \dfrac{c_{m-1}}{c_0} 
\dfrac{A_m}{A_p} 
\dfrac{\alpha_{i,m}}{\alpha_0} 
\dfrac{Q_{i,m}}{Q_{i,\mathrm{max}}} \, .
\end{equation}
In this expression, the first term on the right-hand side represents the active charging site density on the surface before the $m$th impact.
The second term, $A_m/A_p$, denotes the ratio of the contact area to the total particle surface area, $A_p = 4 \pi r^2$.
The third and fourth terms capture the fraction of charge carriers transferred relative to those available in the contact region.

Summing up \cref{eq:dc} over $M-1$ impacts gives the charging site ratio between $t_{M-1}$ and $t_M$, i.e., right before the Mth contact,
\begin{equation}
\label{eq:ctau}
\begin{split}
\dfrac{c}{c_0} (t_{M-1}<&t\le t_M)= 
1 - \sum_{m=1}^{M-1}  
\dfrac{c_{m-1}}{c_0} 
\dfrac{A_m}{A_p} 
\dfrac{\alpha_{i,m}}{\alpha_0} 
\dfrac{Q_{i,m}}{Q_{i,\mathrm{max}}} \\
+ &\sum_{m=1}^{M} 
\left( 1 - \dfrac{c_{m-1}}{c_0} \right)
\dfrac{t_m - t_{m-1}}{\tau_r} \, ,
\end{split}
\end{equation}
with $0 \le c/c_0 \le 1$.
The first summation accounts for depletion over successive impacts, as described by \cref{eq:dc}.
The second term models the slow regeneration of charging sites, which occurs on a characteristic timescale $\tau_r$.
For insulating particles, the timescale $\tau_r$ is typically much larger than the time between contacts.
Thus, regeneration of charge carriers can be neglected, and \cref{eq:ctau} simplifies to
\begin{equation}
\label{eq:c}
\dfrac{c}{c_0} (t_{M-1}<t\le t_M)= 
1 - \sum_{m=1}^{M-1}  
\dfrac{c_{m-1}}{c_0} 
\dfrac{A_m}{A_p} 
\dfrac{\alpha_{i,m}}{\alpha_0} 
\dfrac{Q_{i,m}}{Q_{i,\mathrm{max}}} \, .
\end{equation}
with $0 \le c/c_0 \le 1$.

Assuming $k_e = 1$ and $\chi_p \gg \chi_w$, the contact area ratio $A_m/A_p$ becomes, for particle-wall impacts,
\begin{subequations}
\begin{equation}
\label{eq:awp}
\dfrac{A_m}{A_p} = \dfrac{A_w}{A_p} = \beta_\mathrm{c} \left( \dfrac{5}{128} \pi \rho_\mathrm{p} v_{\mathrm{n},i}^2 \chi_p \right)^{2/5} \, ,
\end{equation}
and for particle-particle interactions,
\begin{equation}
\label{eq:aip}
\dfrac{A_m}{A_p} = \dfrac{A_{ij}}{A_p} = \left( \dfrac{5}{128} \pi \rho_\mathrm{p} v_{\mathrm{n},i}^2 \dfrac{r_j^5}{(r_i + r_j)^2(r_i^3 + r_j^3)} \chi_p \right)^{2/5} \, .
\end{equation}
\end{subequations}

Substituting \cref{eq:awp,eq:aip} into \cref{eq:c} provides the depletion of the charging site density necessary for evaluating \cref{eq:N}.

\subsection{Summary of the SSM}
\label{sec:summary}

To restate for clarity, the key relations of the model are \cref{eq:muw-scale2,eq:mui-scale2,eq:sigmai-scale,eq:gammai-scale,eq:N}.
These relations scale the statistical parameters of charge transfer from an experimentally well-characterized reference impact to a wide range of particle-wall and particle-particle contact conditions occurring in powder flows.

To illustrate the model’s behavior, \cref{fig:spc} shows representative results for generic input parameters (Type~B1).
These plots enable a qualitative discussion of the model response, while \cref{sec:cfd} provides quantitative results for specific conditions.

\Cref{fig:spca} depicts statistical distributions of impact charge for uncharged particles ($\alpha_i/\alpha_0 = \alpha_w/\alpha_0 = 1$), with decreasing charging site density on the particle surface.
As $c/c_0$ decreases, the distributions narrow, and their means shift leftward.
For high $c/c_0$, the distribution exhibits a tail extending into the positive domain.
As the charging sites deplete ($c / c_0 < 0.75$ in this example), the impact charge becomes always negative.
The sign of the impact charge reflects the relative magnitude of charge transfer:
it is positive when the particle-to-wall transfer dominates and negative when the wall-to-particle transfer is larger.
In the limit $c/c_0 \rightarrow 0$, all charging sites on the particle are depleted, and the impact charge converges to a deterministic value determined solely by the wall’s contribution.
Then, the distribution collapses into a vertical line.

\Cref{fig:spcb} shows impact charge distributions for particles with constant $c/c_0=1$ and $\alpha_i/\alpha_0=1$, but decreasing $\alpha_w/\alpha_0$, representing increasing particle charge.
As $\alpha_w/\alpha_0$ decreases, the peaks of the distributions shift to the right because the charge transfer from the wall reduces.
But the charge transfer from the particle to the wall is unaffected by $\alpha_w/\alpha_0$, thus, the widths of the distributions remain constant.
Unlike in \cref{fig:spca}, the distributions remain bipolar until $\alpha_w/\alpha_0$ asymptotically approaches zero.
At saturation, the electric field prevents further net charge transfer from the wall to the particle, and the impact charge is always positive.

\Cref{fig:spcc,fig:spcd} present the evolution of the mean and standard deviation of the impact charge as functions of $\alpha_w/\alpha_0$ and $c/c_0$.
Particles traverse these surfaces over time, moving from the upper right corners (uncharged, carrier-rich) to the lower left corners (saturation-charged, carrier-depleted).
Although the initial and final states may be identical, the path depends on the particle's impact history.

\section{Reference impact experiment}
\label{sec:reference-impact}

\Cref{sec:summary,fig:spc} demonstrate that the statistical parameters of particle-wall charging vary strongly depending on $\alpha/\alpha_0$, $c/c_0$, and other parameters.
Such detailed data have not been reported in the literature, not even in carefully controlled single- and multi-particle charging experiments.
Therefore, we developed a dedicated test rig to measure the statistical input required by the SSM under precisely controlled conditions.
Also, we used this test-rig to test the SSM scaling laws.

A key requirement for the SSM is a clear definition of the initial and impact conditions of the reference impact.
In particular, the conditions that change from impact to impact in the simulations, the particle charge before the impact ($Q$), the contact surface area ($A$), the charging site density ($c$), and the ratio of active charging sites ($\alpha$), must be well controlled.
However, in existing experiments, most of these parameters (especially $c$ and $\alpha$) are not known or not controlled ($Q$)~\citep{Chow18,Mat03,Gro21c,Gro23b}.
To address this lack, in our experiment, contact-free levitation allows to de-charge and reset the particle before impact, ensuring that each impact is a first impact of a fresh particle.
Experiments based on repeated impacts of the same particles (e.g.,~\citep{Grosj23a}) cannot provide such data, since $c$ progressively depletes.

The SSM requires the first three moments of the reference impact's charge distribution ($\mu_0$, $\sigma_0$, $\gamma_0$) as well as its minimum impact charge ($\Delta Q_{0,\mathrm{min}}$).
These quantities are rarely reported in the literature.
For example, \citet{Lee18} provide mean values over many impacts, but not the full statistical descriptors necessary for our model.

Furthermore, the SSM does not yet include scaling with respect to other environmental or material parameters such as humidity, pressure, temperature, particle material, or shape.
Consequently, the reference experiment must be performed under the same ambient conditions as the CFD simulation.
This restriction precludes the use of vacuum experiments (e.g.,~\citep{Cart19}).
To meet these requirements, our rig is installed in an extremely well-controlled climate chamber, where all relevant environmental parameters can be maintained constant.

Thus, through our experimental approach, the measurement of the required statistical parameters and the validation of the SSM scaling laws can be carried out without any parameter tuning or manually adjusting model constants..

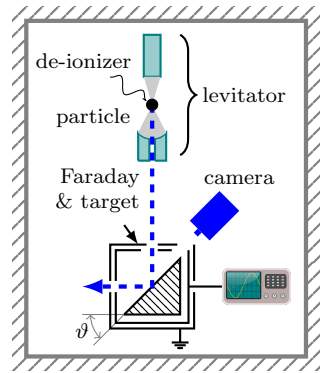
\begin{figure}[tb]
\centering
\begin{center}
\begin{tikzpicture}[scale=0.37,font=\footnotesize]
\coordinate (p) at (1,7.5); 
\path[pattern={Lines[angle=45,distance={6pt/sqrt(2)}]},pattern color=gray, thick] (-4,-2) rectangle (7,11);
\draw [ultra thick,gray,fill=white] (-3.5,-1.5) rectangle (6.5,10.5);

\draw [gray] (-1.1,-1.1) -- (0,0) -- (-1.5,0) node [below,black] {$\vartheta$};
\draw [<->,gray] (-1.2,0) arc [radius=1.2, start angle=180, end angle=225];
\draw [thick,pattern=north west lines] (0,0) -- (2,0) -- (2,2) -- (0,0); 
\draw [thick] (-.5,.8) -- (-.5,-.5) -- (2.5,-.5) -- (2.5,.8); 
\draw [thick] (2.5,1.2) -- (2.5,2.5); 
\draw [thick] (1.9,2.5) -- (1.2,2.5);
\draw [thick] (.8,2.5) -- (-.5,2.5) -- (-.5,1.2);
\draw [thick] (2,-1) -- (2,-.5);
\draw [thick] (1.7,-1) -- (2.3,-1);
\draw [thick] (1.8,-1.1) -- (2.2,-1.1);
\draw [thick] (1.9,-1.2) -- (2.1,-1.2);
\draw [thick] (-.3,.8) -- (-.3,-.3) -- (2.3,-.3) -- (2.3,2.3); 
\draw [thick] (1.7,2.3) -- (1.2,2.3);
\draw [thick] (.6,2.3) -- (-.3,2.3) -- (-.3,1.2);
\draw [thick] (2.3,1) -- (3.5,1); 
\node [right] at (3.2,1) {\includegraphics[trim=0mm 0mm 0mm 0mm,clip=true,width=0.05\textwidth]{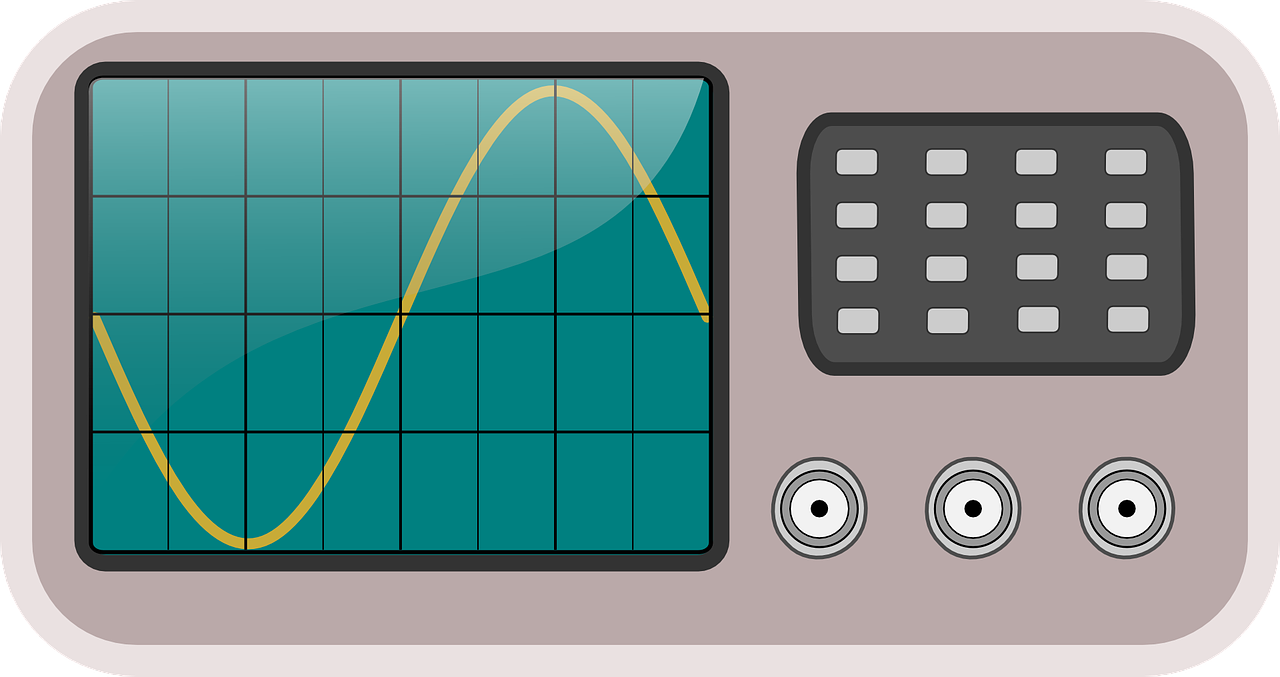}};
\draw [thick,teal,fill=teal!40] (.7,8.5) rectangle (1.3,10); 
\draw [thick,teal,fill=teal!40] (1.1,5.5) -- (1.5,5.5) -- (1.5,6.5) to [out=225,in=360] (1.1,6.3) -- (1.1,5.5);
\draw [thick,teal,fill=teal!40] (0.5,5.5) -- (0.9,5.5) -- (0.9,6.3) to [out=180,in=315] (0.5,6.5) -- (0.5,5.5);
\path [fill=gray!40] (p) -- (1.5,6.5) to [out=225,in=315] (.5,6.5) -- (p);
\path [fill=gray!40] (p) -- (1.3,8.5) -- (.7,8.5) -- (p);
\draw [->,>=latex,ultra thick,dashed,blue] (p) -- (1,1) -- (-1.5,1); 
\draw [fill=black,black] (p) circle [radius=.2] node [left,xshift=-5,yshift=-5] {particle};
\draw [shorten >=0.15cm,style={decorate,decoration=snake} ] (-.5,8.5) node [align=right,above,xshift=-10] {de-ionizer} -- (p);
\draw [thick,->,>=latex] (0,3.2) node [align=right,yshift=13,xshift=-10] {Faraday\\ \& target} -- (0.5,2.5);
\draw [thick,decorate,decoration={brace,amplitude=6pt,mirror}] (2,5.7) -- (2,10) node[midway,right,xshift=5]{levitator};
\path [fill=blue,rotate around={-45:(2.7,3)}] (2.2,3) rectangle (3.2,4.5) node [above,yshift=7] {camera};
\path [fill=blue,rotate around={-45:(2.7,3)}] (2.6,2.8) rectangle (2.8,3);
\end{tikzpicture}
\end{center}
\caption{%
(a) Schematic of the particle impact charging apparatus (not to scale). 
For each reference impact condition, fresh, uncharged particles impact the target under identical conditions.}
\label{fig:reference-impact-sketch}
\end{figure}

\begin{figure*}[tb]
\centering
\begin{center}
\subfloat[Particle: PS, Target: aluminum]{\includegraphics[trim=0mm 0mm 0mm 0mm,clip=true,height=0.27\textwidth]{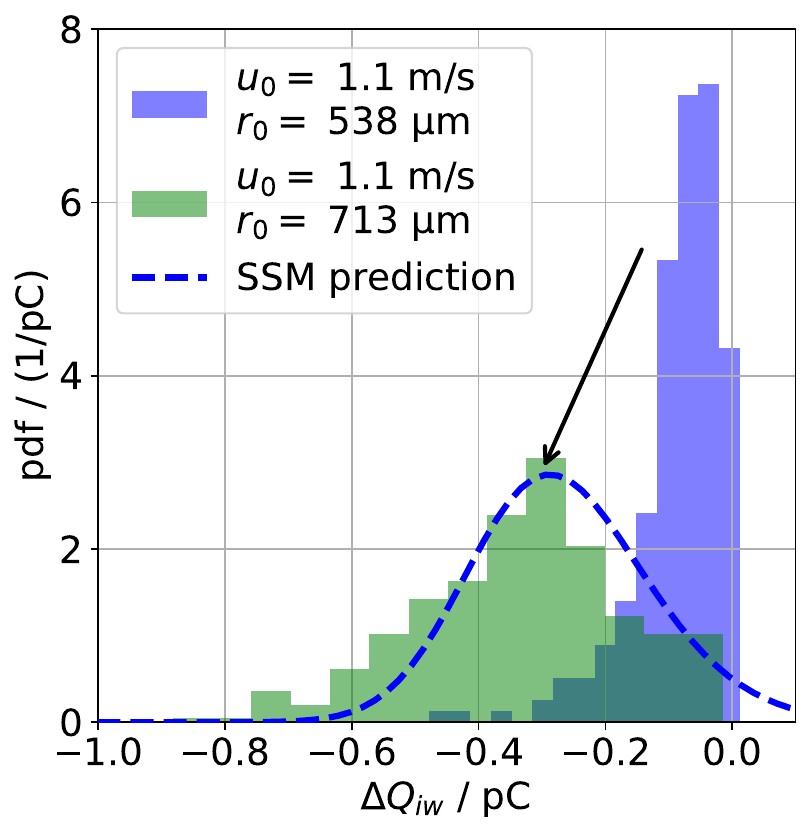}\label{fig:binom-12-3}}\quad
\qquad
\subfloat[Particle: PMMA, Target: aluminum]{\includegraphics[trim=0mm 0mm 0mm 0mm,clip=true,height=0.27\textwidth]{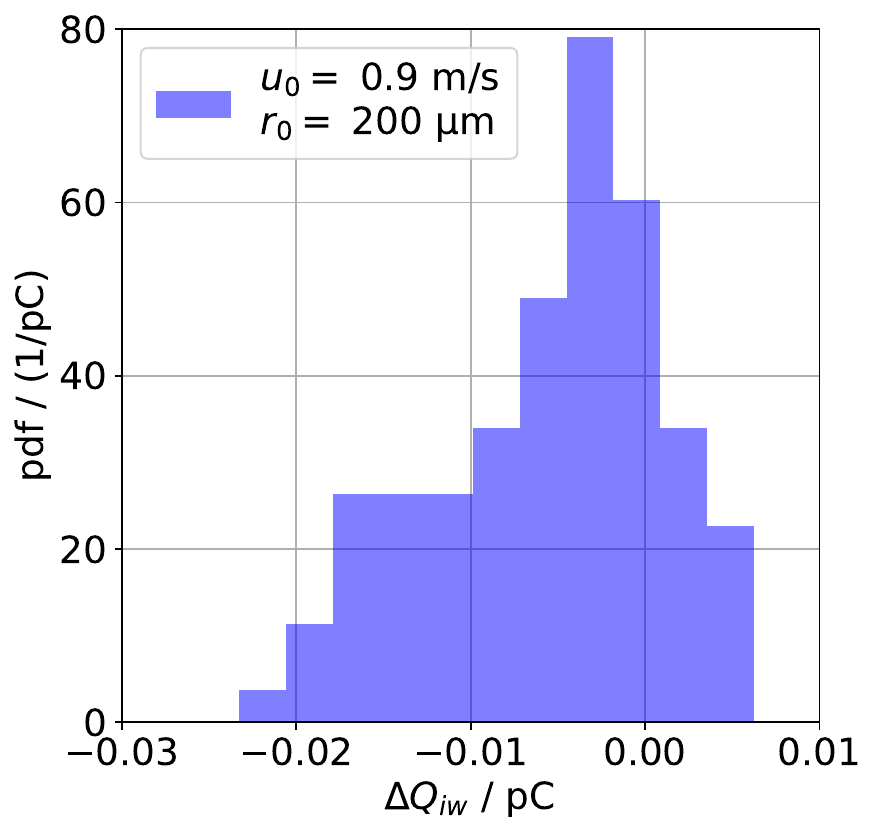}\label{fig:binom-2}}
\qquad
\subfloat[Particle: PMMA, $r_0=750~\upmu$m, $v_0=1.7$~m/s]{\includegraphics[trim=0mm 0mm 0mm 0mm,clip=true,height=0.27\textwidth]{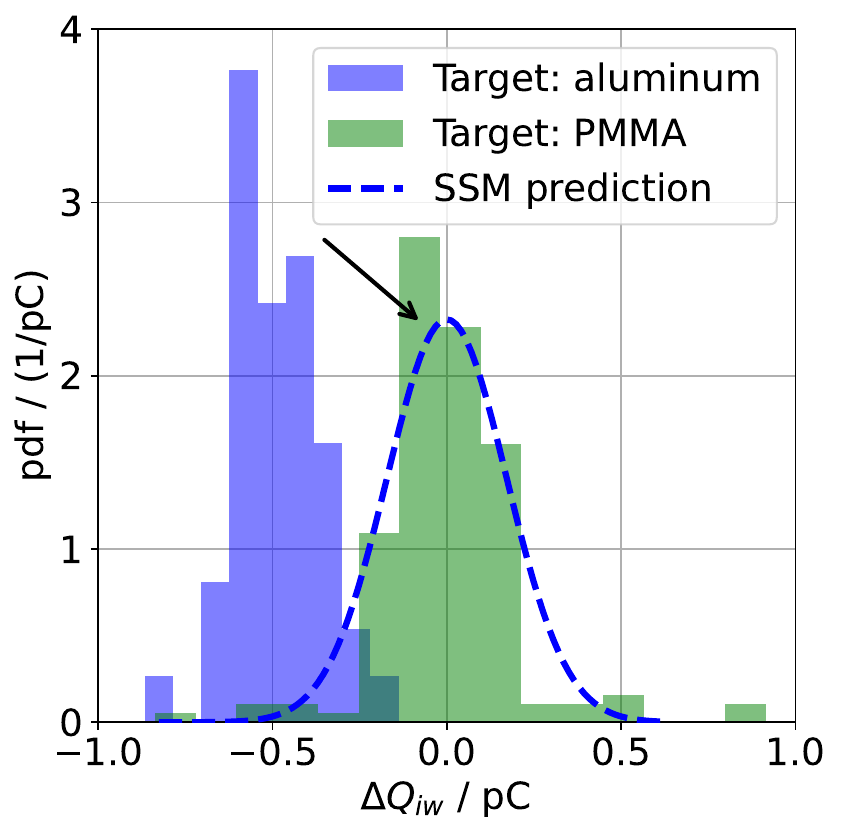}\label{fig:binom-3}}
\end{center}
\caption{%
Results of the reference impact experiment.
The histograms display the raw measured charge distributions;
the dashed curves the predictions by SSM.
(a)~PS particles of two different sizes. 
The SSM predicts the distribution of the larger particles by scaling the statistical parameters obtained from the smaller particles.
(b)~Reference impact results for PMMA particles.
(c)~Identical PMMA particles impacting an aluminum and a PMMA target.
The SSM predicts the distribution of the same-material impacts by scaling the statistical parameters obtained from the different material impacts.}
\label{fig:reference-impact}
\end{figure*}

\begin{table*}[tb]
\caption{\label{tab:reference-impact} Reference impact experiments.}
\begin{ruledtabular}
\begin{tabular}{lccccc}
Particle                         & PS      & PS      & PMMA     & PMMA     & PMMA \\
Target                           & aluminum& aluminum& aluminum & aluminum & PMMA \\
$r_0$ ($\upmu$m)                 & 538     & 713     & 200      & 750      & 750 \\ 
$v_0$ (m/s)                      & 1.1     & 1.1     & 0.9      & 1.7      & 1.7 \\
Impacts                          & 241     & 317     & 100      & 46       & 165 \\
\hline
$\mu_0$ (fC)                     & -68.5   & -306.6  & -4.1     & -444.9   & 56.7 \\
$\sigma_0$ (fC)                  & 72.3    & 183.5   & 6.5      & 121.3    & 191.9 \\
$\gamma_0$ (-)                   & -0.56   & -0.45   & -0.19    & -0.03    & 0.02 \\
$\Delta Q_{0,\mathrm{min}}$ (fC) & -462.0  & -1842.0 & -21.9    & -825.6   & -778.8 \\
\end{tabular}
\end{ruledtabular}
\end{table*}

\subsection{Statistical parameters: $\mu_0$, $\sigma_0$, $\gamma_0$, $\Delta Q_{0,\mathrm{min}}$}

The SSM derived in the previous section requires four input parameters (\cref{fig:variables}) for a defined reference impact:
$\mu_0$, $\sigma_0$, $\gamma_0$, and $\Delta Q_{0,\mathrm{min}}$.
For the skewness, we found it more accurate to use the alternative expression \cref{eq:gamma-alt} with $p=(\mu_0-\Delta Q_{0,\mathrm{min}})/(\Delta Q_{0,\mathrm{max}}-\Delta Q_{0,\mathrm{min}})$ and with $\Delta Q_{0,\mathrm{max}}$ being the maximal impact charge of the reference impact.
As shown in Appendix~\ref{sec:qimax}, this expression is equivalent to the skew of the raw data for converged statistical distributions.

To provide accurate and material-specific input data, we developed an apparatus to measure these quantities.
This experiment enables precise control over initial and boundary conditions, minimizing scattering in the results.
The setup, sketched in \cref{fig:reference-impact-sketch}, achieves a measurement uncertainty of less than 1\,fC for the impact charge of a single particle.

\begin{figure}[tb]
\begin{center}
\includegraphics[trim=0mm 0mm 0mm 0mm,clip=true,width=0.35\textwidth]{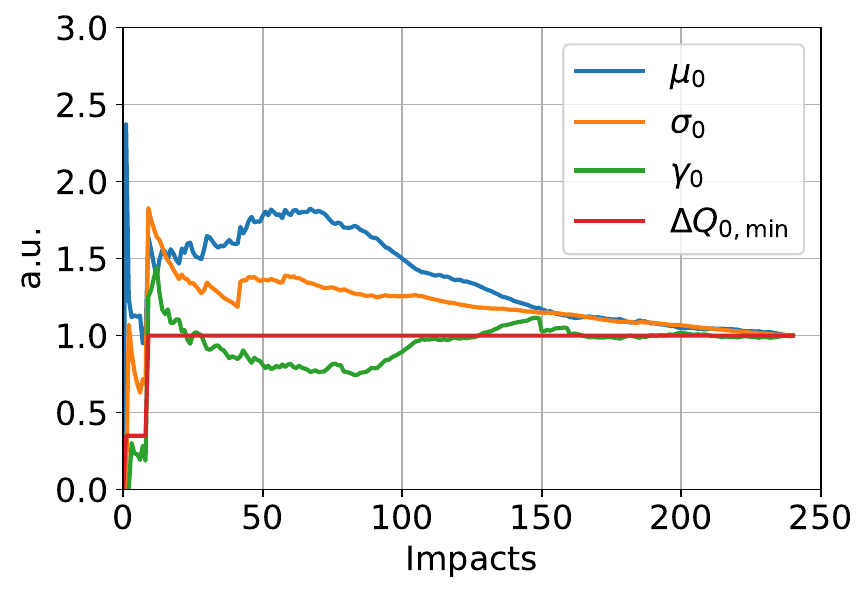}
\end{center}
\caption{%
Convergence of the statistical parameters of the reference impact of PS particles with $r=538$~µm.
Each parameter is plotted relative to its final value after 241 impacts.}
\label{fig:reference-impact-convergence}
\end{figure}

The apparatus integrates and extends two established techniques, the experimental setup by \citet{Mat03} and acoustic levitation.
In comparable facilities, scattering often arises from unknown initial charge distributions on the particle surface.
To mitigate this uncertainty, particles are initially levitated in an acoustic trap.
This contact-free handling prevents unwanted triboelectric charging and allows the neutralization of surface charges using an ionization needle.
Additionally, acoustic streaming clears ions from the vicinity of the particle, further reducing the unintended charge.
Upon deactivating the levitator, the particle falls freely through a hole in the reflector towards the target.

All measurements were conducted in a climate-controlled chamber, where ambient temperature and relative humidity were maintained at 25\,\textdegree C and 10\%, respectively.
The drop height of 10~cm to 15~cm allows particles of $r=200$\,\textmu m to reach normal impact velocities of approximately 0.9\,m/s due to gravity.
This velocity is representative of typical wall-normal impact velocities in flows parallel to walls, which are generally below 10\% of the streamwise gas velocity.

The particle then enters a Faraday cage, impacts a conductive target (15\,mm\,$\times$\,15\,mm, $\vartheta = 35$\textdegree), and exits the cage.
A high-speed CMOS camera with an integrated light source observes the target through a small optical window.
These images are used to determine the particle's impact velocity.
A digital oscilloscope, sampling at 25\,\textmu s to 500\,\textmu s intervals, records the initial and final charge of the particle.

The reference impact conditions and measured statistical parameters are listed in \cref{tab:reference-impact} for two powders, polystyrene (PS) and polymethyl methacrylate (PMMA), interacting with an aluminum wall.
Experimental charge distributions and model fits are shown in \cref{fig:reference-impact}.

\Cref{fig:reference-impact-convergence} plots the sensitivity of the statistical parameters with the number of impacts for $r=538$~µm sized PS particles.
According to the figure, the statistical errors drop below 20\% after 150 impacts.
For each reference experiment in \cref{fig:reference-impact}, we measured impacts until the statistical parameters converged.

\subsection{Test of the scaling laws}
\label{sec:test-scaling}

For each material pair, the model requires only a single reference condition.
Nonetheless, we conducted two reference experiments for PS to evaluate the accuracy of the model’s scaling.
These cases differ solely in particle size.
As shown in \cref{fig:binom-12-3}, the SSM successfully predicts the impact statistics of the larger particles based on the smaller-particle reference.
The absolute error of the scaling by SSM compared to the measured distribution is 0.040~pC for the mean, 0.041~pC for the standard deviation, and 0.73 for the skewness.

These deviations stem from several sources, one being the statistical error due to a finite number of measurements.
For a normally distributed sample, the standard error of the skewness ($\approx(M/6)^{-1/2}$) drops the slower with the number of observations, $M$, compared to the standard errors of the mean ($=M^{-1/2}$) and standard deviation ($\approx(2M)^{-1/2}$).
The error in $\Delta Q_{0,\mathrm{min}}$ is binary;
either the dataset captures the minimal impact charge or not, where the probability of capturing the minimal impact charge is $1-(1-(1-p)^{N_0})^M$.
Thus, the statistical uncertainties associated with $\gamma_0$ and $\Delta Q_{0,\mathrm{min}}$ are the most difficult to constrain.

Further, we test the assumption that the charge-transfer distribution for same-material contacts can be obtained by scaling the distribution measured for different-material contacts.
To this end, we conducted two reference experiments with PMMA particles, differing only in the target material, which was aluminum in one case and PMMA in the other.
As shown in \cref{fig:binom-3}, for the investigated PMMA particles, the SSM successfully predicts the statistical characteristics of same-material impacts based solely on the statistics of different-material charging.
The absolute errors between the SSM-scaled predictions and the measured distributions are 0.057~pC for the mean, 0.020~pC for the standard deviation, and 0.02 for the skewness.

After having established the statistical parameter's scaling accuracy (\cref{fig:reference-impact}) and convergence (\cref{fig:reference-impact-convergence}), we investigate the sensitivity of the model to these parameters in \cref{sec:err}.

\section{CFD simulations}
\label{sec:cfd}

After testing the scaling laws in \cref{sec:test-scaling}, we assess whether the SSM can predict the three charging patterns highlighted in the Introduction: \emph{variable impact charge}, \emph{charge reversal}, and \emph{size-dependent bipolar charging}.
To do so, we conducted CFD simulations with thousands of PS and PMMA particles.
In these simulations, the SSM predicts these charging patterns for surfaces up to 400~million times larger than the mosaic model by \citet{Grosj23}, enabling the computation of complete powders. 

The particles were released in canonical geometries and flow conditions, specifically, wall-bounded turbulent flows and periodic cubic boxes.
We chose wall-bounded turbulent flows because they mimic the conditions in pneumatic conveyors, where particles acquire the highest charges in industry~\citep{Kli18}, as well as in other powder-handling devices and wall-bounded natural flows.

Consistent with the reference impact measurements (\cref{sec:reference-impact}), all walls in the simulations are composed of aluminum, the ambient temperature is 25\,\textdegree C, and the relative humidity is 10\%.
\Cref{tab:cfd} provides an overview of the parameters of all simulations. 

We implemented the SSM (\cref{sec:spc}) into the open-source CFD tool \citet{pafiX}.
The code employs a coupled Eulerian-Lagrangian scheme in which particle-particle collisions are modeled using a hard-sphere approach.
The carrier gas flow is resolved by Direct Numerical Simulation (DNS) of the incompressible Navier-Stokes equations, capturing all turbulent scales.

Individual particles are tracked in a Lagrangian reference frame, and their trajectories are computed by solving Newton’s second law, accounting for drag, lift, electrostatic, gravitational, and collisional forces.
A hybrid approach~\citep{Gro17e} computes the electrostatic field, obtaining the near-field of neighboring particles from Coulomb's law and the far-field from Gauss’s law.
Collisional forces account for particle-particle and particle-wall interactions.
Further details on the solver setup, numerical methods, and validation are given in \cref{sec:math}.

To evaluate \emph{variable impact charge} and \emph{charge reversal}, we simulated 300\,000 particles, having a total surface area of up to~0.15\,m$^2$, in a wall-bounded turbulent flow (simulations~1--4; \cref{tab:cfd}).
The aim to capture charge reversal was one of the reasons for testing the SSM within an Euler-Lagrange framework, which, unlike Euler-Euler methods, resolves the time histories of individual particles.

The domain geometry corresponds to a downward-oriented square duct with a side length of 5\,cm.
We simulate a duct section with periodic boundary conditions in the streamwise direction, driving the flow by a constant pressure gradient and gravitational acceleration.
Based on the duct width, the flow has a bulk Reynolds number of approximately~5400 and a friction Reynolds number of~360.
Each simulation begins with electrostatics disabled to allow the gas and particle flow to reach a statistically stationary state.
Once the flow is fully developed, the SSM is activated.

\begin{figure}[tb]
\begin{center}
\includegraphics[trim=0mm 0mm 0mm 0mm,clip=true,width=0.40\textwidth]{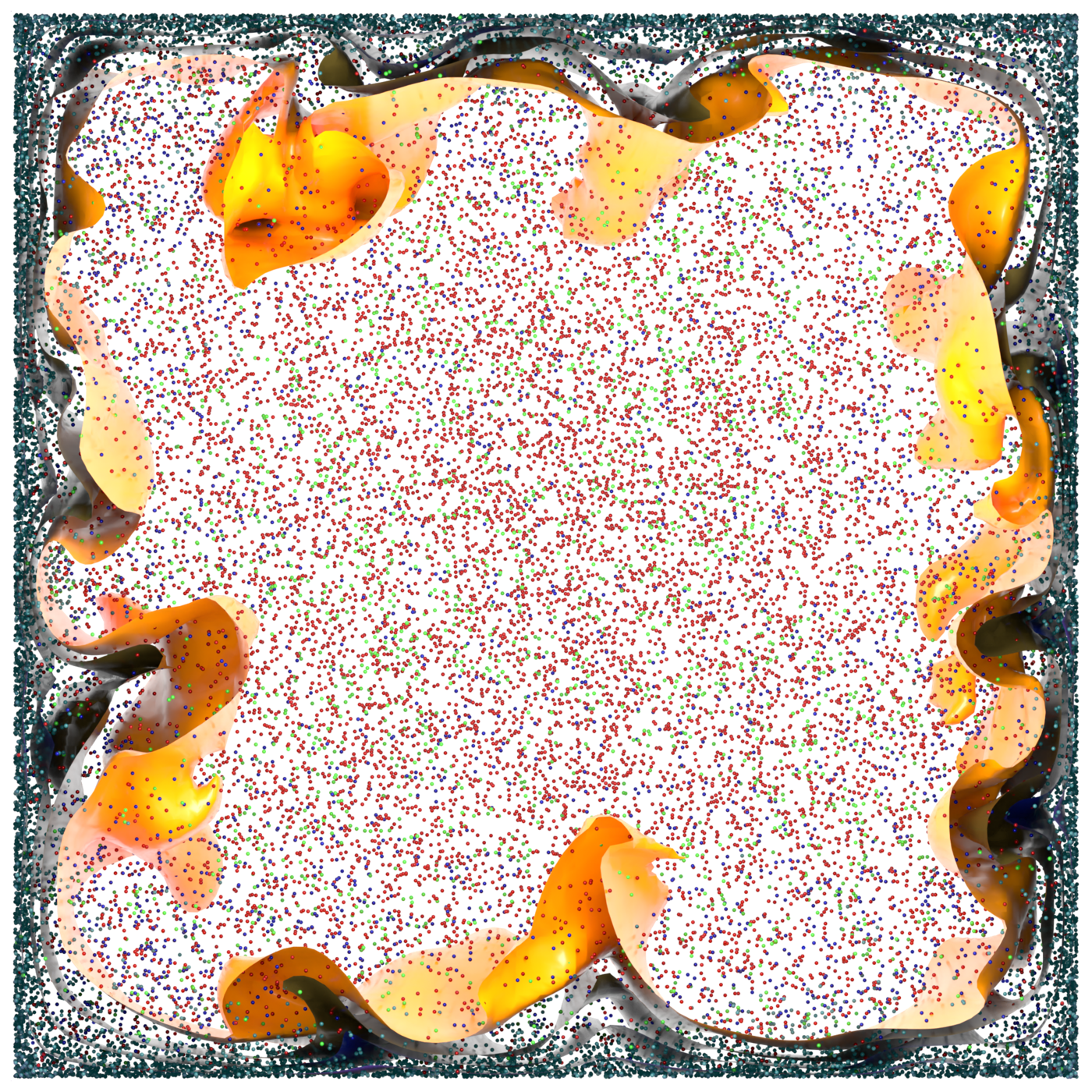}
\end{center}
\caption{%
Snapshot of 200~µm sized PMMA particles in a wall-bounded turbulent flow (simulation~2; \cref{tab:cfd}).
The snapshot depicts the cross-section and 10\% of the domain in the downstream direction at the time instance when $\overline{c/c_0}=0.5$.
The isosurfaces visualize constant streamwise velocity magnitude of the gas (0.38~m/s, 0.75~m/s, 1.13~m/s, and 1.50~m/s).
Cyan particles contacted a wall, and green particles contacted another particle within the last millisecond.
Particles that did not collide recently are colored by their polarity:
red indicates positive, blue negative charge.}
\label{fig:flow}
\end{figure}

\begin{figure*}[tb]
\begin{center}
\subfloat[PS and PMMA, 200~µm]{\includegraphics[trim=0mm 0mm 0mm 0mm,clip=true,height=0.22\textwidth]{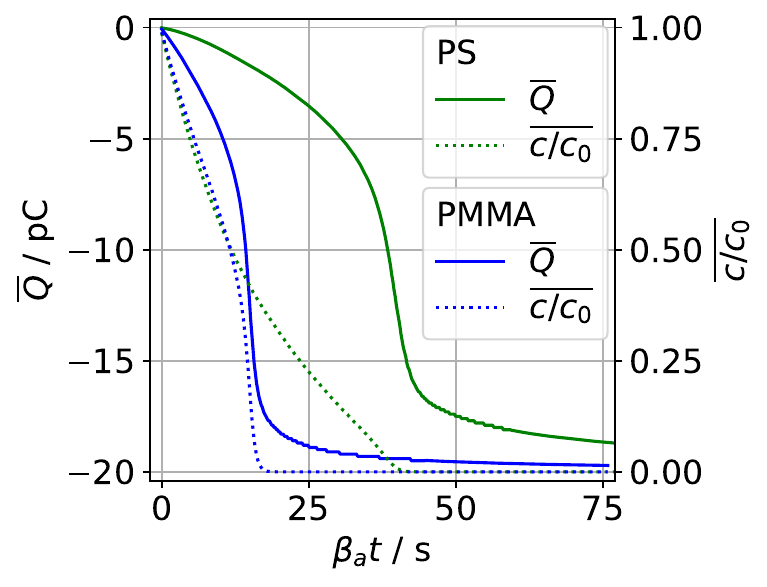}\label{fig:q-c-t-200}}\qquad
\subfloat[PS, 200~µm]{\includegraphics[trim=0mm 0mm 0mm 0mm,clip=true,height=0.22\textwidth]{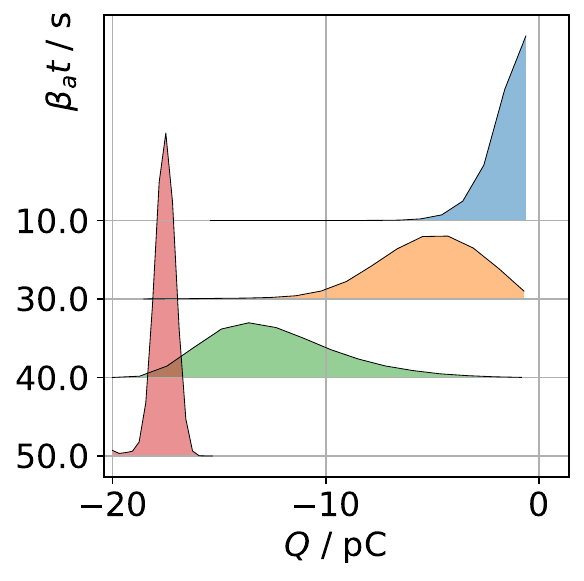}\label{fig:distribution-ps200}}\qquad
\subfloat[PMMA, 200~µm]{\includegraphics[trim=0mm 0mm 0mm 0mm,clip=true,height=0.22\textwidth]{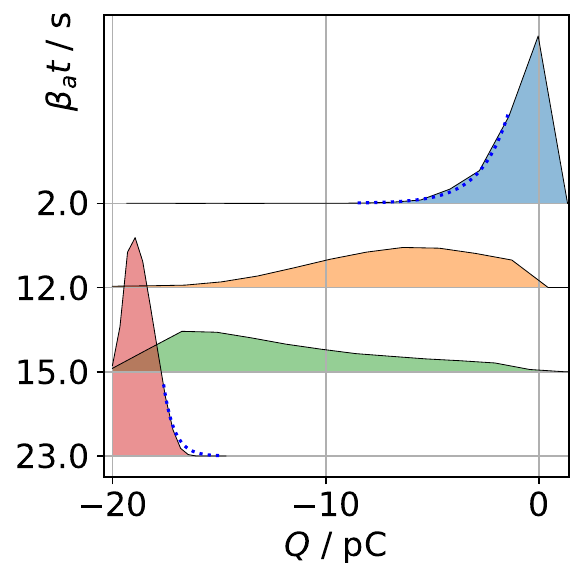}\label{fig:distribution-pmma200}}
\end{center}
\caption{%
(a) Temporal evolution of the particle charge and charging site density for 200~\textmu m PS and PMMA particles transported in wall-bounded turbulent flow.
(b) Charge distribution of PS particles at selected time instances.
(c) Charge distribution of PMMA particles at selected time instances.
The blue dotted lines in (c) indicates a logarithmic fit to the distributions at 2.0~s and 23.0~s.}
\label{fig:sim-overview}
\end{figure*}

To assess \emph{size-dependent bipolar charging}, we simulated 62\,500 particles in a cubic domain of side length 5\,cm (simulations~5 and~6; \cref{tab:cfd}).
To isolate the effect of particle-particle collisions, we employed periodic boundary conditions on all sides, thereby eliminating wall contacts.
Further, only collisional and electrostatic forces act on the particles; fluid forces (drag and lift) and gravity are omitted.
Particles are initialized at random positions with velocities of 5\,m/s in random directions.

Particles are either PS, with a density of $\rho_\mathrm{p} = 1000$~kg\,m$^{-3}$, or PMMA, with a density of $\rho_\mathrm{p} = 1150$~kg\,m$^{-3}$.
Due to their similar elastic properties, both materials are assigned the same elasticity parameter, $\chi_\mathrm{p} = 2.9 \times 10^{-10}$~s$^2$\,m\,kg$^{-1}$~\citep{Dub07,Dom86}.
The aluminum walls are modeled with an elasticity parameter of $\chi_\mathrm{w} = 1.25 \times 10^{-11}$~s$^2$\,m\,kg$^{-1}$~\citep{Dom86}.
In all simulations, the coefficient of restitution is unity.
We report results for both monodisperse systems (with particle diameters of 100~µm or 200~µm) and bidisperse mixtures containing both sizes.

\Cref{fig:flow} shows a snapshot from simulation~1 (\cref{tab:cfd}) of 300\,000 PMMA particles (monodisperse, $r=$~200~µm) in a wall-bounded turbulent flow.
The figure shows 10\% of the total flow domain and particle number.
The isosurfaces of the streamwise velocity visualize the complex fluid structures.
The particles are colored by their instantaneous net charge or collision history.
Particles that have collided with a wall within the last millisecond are rendered in cyan, whereas those that recently underwent a particle-particle collision appear in green.
This visualization highlights the dynamic interplay between turbulent transport and collisional charging.

As particles interact with walls and each other, their charge evolves in time according to the SSM.
In the depicted case, 300\,000 particles collide on average 1.5 million times per second with a wall and 1.8 million times per second with each other.
Despite its particle-wise application at each collision, the SSM remains computationally efficient, accounting for less than 0.01\% of the total simulation runtime.
This efficiency enables large-scale simulations with fully resolved charge evolution at the particle scale, even in turbulent flow.

\begin{figure*}[tb]
\begin{center}
\subfloat[]{\includegraphics[trim=0mm 0mm 0mm 0mm,clip=true,height=.22\textwidth]{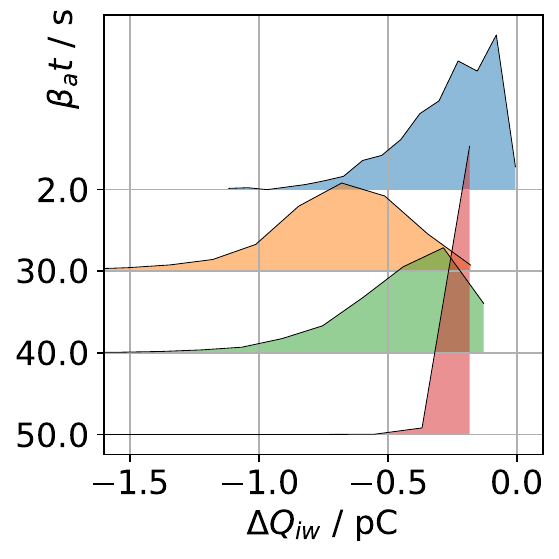}\label{fig:dqpw-distribution-ps200}}\qquad
\subfloat[]{\includegraphics[trim=0mm 0mm 0mm 0mm,clip=true,height=.22\textwidth]{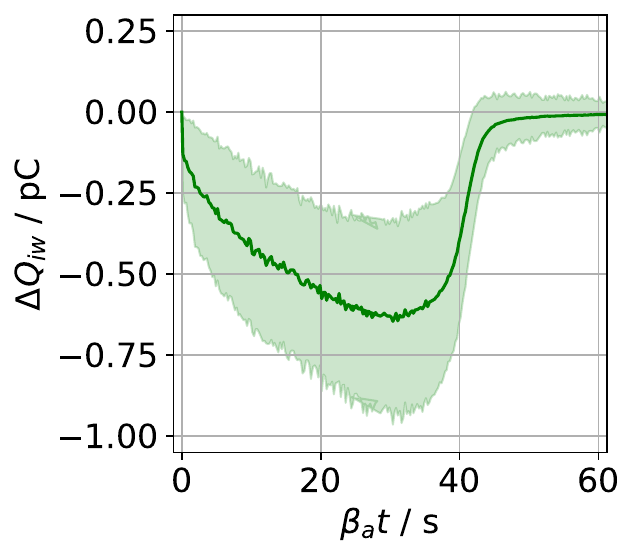}\label{fig:dqpw-t-ps200}}\qquad
\subfloat[]{\includegraphics[trim=0mm 0mm 0mm 0mm,clip=true,height=.22\textwidth]{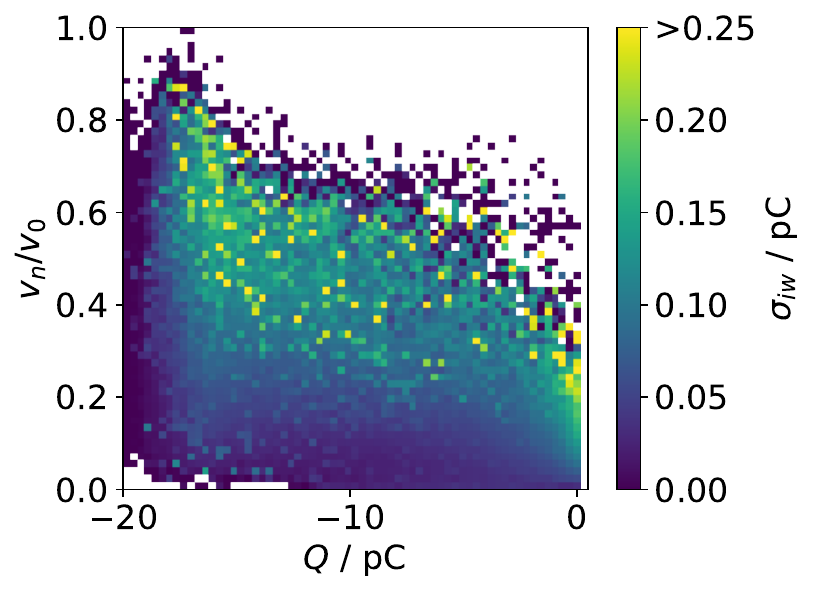}\label{fig:dqpw-heat-ps200}}
\end{center}
\caption[]{%
(a) Variable impact charge of 200~µm sized PS particles predicted by the SSM at selected time instances, consistent with experimental observations (cf.~\cref{fig:condenserb}).
(b) Temporal evolution of the mean and standard deviation of the impact charge reflects changing impact conditions and particle properties over the course of the simulation.
(c) For constant impact velocity and initial charge, the variation in impact charge arises from the stochastic particle-to-surface charge transfer as described by \cref{eq:skew-normal}.}
\label{fig:variable-impact-charge-ps}
\end{figure*}

\Cref{fig:q-c-t-200} shows the temporal evolution of the average particle charge and the charging site density for 200~\textmu m PS and PMMA particles (simulations 1 and 2; \cref{tab:cfd}).
PMMA charges faster than PS.
In both cases, the charging sites fully deplete over time, and the particles reach their saturation charges.
For PMMA, the evolution of the normalized charge closely follows the depletion of the charging sites.
In contrast, for PS, the increase in particle charge lags behind the depletion of charging sites.
This delay arises because, for PS, the contribution of particle-to-wall charge transfer is large, which reduces the net charging.
Consequently, although PS particles lose their charging sites, they initially accumulate less charge.
Only after some amount of charging sites has been depleted, after about 35~s, does the net impact charge increase significantly, and PS charges at a rate comparable to that of PMMA.

\Cref{fig:distribution-ps200,fig:distribution-pmma200} present the corresponding charge distributions of the 200~\textmu m PS and PMMA particles at selected time instances.
For both materials, the SSM captures the smooth transition of the distributions from centering around uncharged particles toward peaking near the saturation charge.

The particle charge distributions approximate logarithmic functions at early and late times in the simulations.
In \cref{fig:distribution-pmma200}, the blue dotted curves illustrate the close agreement of the PMMA charge distribution at the beginning and the end of the charging with a logarithmic function of positive slope.
After about 2~s, the left tails of the distributions are logarithmic with a positive slope, and after about 23~s, their right tails are logarithmic with a negative slope.
Thus, a mixture of particles at different stages in their charging history, some having undergone few and some many collisions, would produce charge distributions with two exponential tails.
This behavior demonstrates the SSM's ability to reproduce non-Gaussian charge distributions as observed experimentally~\citep{Mujica23,Haeb18}.

Having established the general features of the charging dynamics, we now examine the three charging patterns highlighted in the Introduction.

\subsection{Variable impact charge}

We first demonstrate by \cref{fig:variable-impact-charge-ps} that the SSM predicts \emph{variable impact charge}, consistent with experimental observations (\cref{fig:condenserb}).
Particles acquire different amounts of charge upon contact with a surface, even under nominally identical impact conditions.
As evident from \cref{eq:skew-normal}, this variability arises from the stochastic charge transfer from the insulating particle to another surface.
The model thus inherently captures the experimentally observed spread in impact charge.

For wall-bounded turbulent flow, \cref{fig:dqpw-distribution-ps200} shows the distributions of impact charge for 200~µm PS particles during wall collisions at selected time instances (simulation~1; cf.~\cref{tab:cfd}).
These distributions evolve over time, reflecting changes in the overall impact charge statistics driven by the depletion of charging sites and increasing particle charge.
As shown in \cref{fig:dqpw-t-ps200}, the initial impact charges are low due to the near balance between particle-to-wall and wall-to-particle charge transfer.

Until $t=32$~s, the magnitude of the impact charge increases because the charging sites deplete, and wall-to-particle transfer dominates more and more.
During this time, the impact charge distributions in \cref{fig:dqpw-distribution-ps200} move toward the left.
After approximately $32$~s, the increasing particle charge reduces the net impact charge due to electrostatic inhibition of further transfer and the impact charge distributions move back toward the right.
At $t=45$~s, particles approach saturation, and the mean impact charge and its standard deviation tend toward zero.

However, even for identical initial charge and impact velocity, the impact charge remains variable, as shown in \cref{fig:dqpw-heat-ps200}.
The variation of the impact charge is the highest for initially uncharged particles at high impact velocities, where the number of active charging sites, i.e., $N_i/N_0$ in \cref{eq:N}, is large.

The SSM, therefore, reproduces the variable impact charge observed in experiments~\citep{Mat03,Mat06c,Grosj23a}.
This behavior supports the hypothesis that the variability arises from the stochastic charge transfer at individual contact sites on insulating surfaces.

The impact charge variability of PMMA particles (simulation~2; cf.~\cref{tab:cfd}) is comparable to that of PS particles, as discussed in \cref{sec:variable-impact-charge-pmma}.

\subsection{Charge reversal}

\begin{figure}[b]
\begin{center}
\subfloat[PS, $r=100$~µm]{\includegraphics[trim=0mm 0mm 0mm 0mm,clip=true,width=0.24\textwidth]{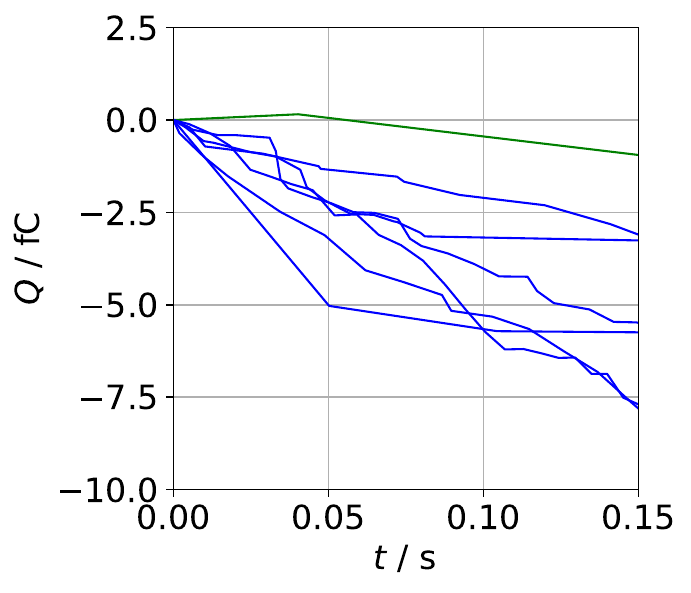}\label{fig:}}~~
\subfloat[PMMA, $r=100$~µm]{\includegraphics[trim=0mm 0mm 0mm 0mm,clip=true,width=0.24\textwidth]{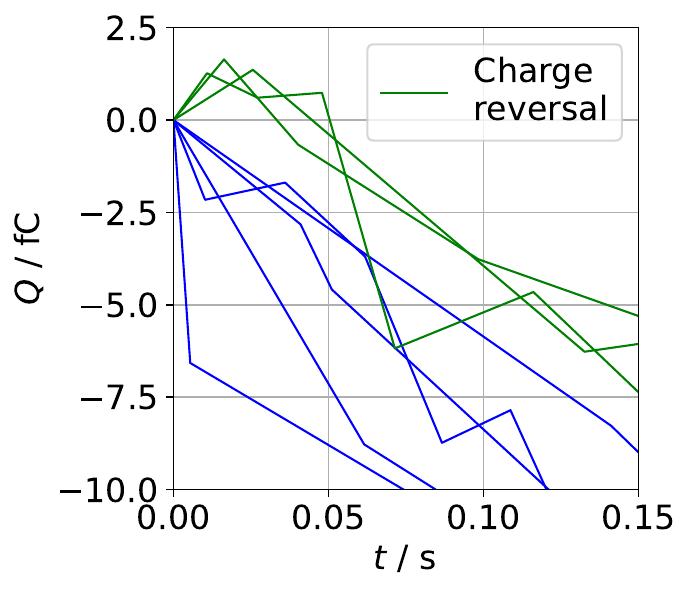}\label{fig:}}
\end{center}
\caption[]{%
Charge reversal during wall contacts of selected particles, as predicted by the SSM and observed in experiments (\cref{fig:condenserc}).
Blue curves represent particles that charge negatively throughout the simulation.
Green curves show particles that undergo charge reversal, first charging positive and then negative, consistent with experimental findings~\citep{Low86b,Shaw28}.
The probability of initially acquiring a positive charge is about five times higher for PMMA particles than for PS particles.}
\label{fig:charge-reversal}
\end{figure}

Second, we demonstrate \emph{charge reversal} by analyzing the particle polarity during successive impacts, as shown in \cref{fig:charge-reversal}.
To this end, we plot the charge histories of selected 100~µm sized PS and PMMA particles in wall-bounded turbulent flow (simulations~3 and~4; cf.~\cref{tab:cfd}).
Particles can undergo both positive and negative charge transfers during collisions with other particles because there is no inherent directionality in same-material charge transfer.
Therefore, to isolate charge reversal at wall contacts with a different material, we neglect particle-particle charging in the simulations shown in \cref{fig:charge-reversal}.

As illustrated, the SSM predicts charge reversal for some particles and not for others, qualitatively consistent with experimental observations (\cref{fig:condenserc}).
In agreement with those experiments, the polarity typically switches from positive to negative after the first or second wall contact.
This outcome corresponds to the hypothesized mechanism underlying charge reversal:
During the first impact, the particle surface contains a full layer of active charging sites.
Depleting this layer can result in a net positive charge transfer from the wall to the particle.
As collisions continue, the active charging sites on the particle surface deplete.
The charging sites being depleted mostly shifts the charge transfer balance so that charge transfer from the wall to the particle dominates subsequent impacts, producing a net negative impact charge.

In wall-bounded turbulent flow, approximately 27.2\% of the 100~µm PMMA particles initially charge positively, compared to only 6.1\% of the PS particles.
These probabilities differ from those in the reference impact statistics (\cref{fig:reference-impact}).
This difference is because the mean charge transfer and its variability scale differently:
The charge transfer from the wall to the particle, $\mu_w$, scales proportionally to the ratio of active charging sites, which, during the first impact, is proportional to the contact area ratio (cf.~\cref{eq:muw}).
Similarly, the mean charge transfer from the particle to the wall, $\mu_i$, scales with the same ratio (cf.~\cref{eq:mui}).
However, the standard deviation of the transfer, $\sigma_i$, scales with the square root of the contact area (cf.~\cref{eq:sigmai}).
Since $\sigma_i$ determines the likelihood that the particle-to-wall transfer exceeds the reverse, the probability of charge reversal is higher for particles with smaller contact areas.
This explains why charge reversal is more prevalent in the 100~µm particles than in the 200~µm particles.

Importantly, the SSM predicts charge reversal solely based on the scaling laws and the stochastic charge transfer statistics of the reference impact.
The model contains no explicit switching mechanism.

\subsection{Size-dependent bipolar charging}

\begin{figure*}[tb]
\begin{center}
\subfloat[PS, bidisperse]{\includegraphics[trim=0mm 0mm 0mm 0mm,clip=true,height=0.22\textwidth]{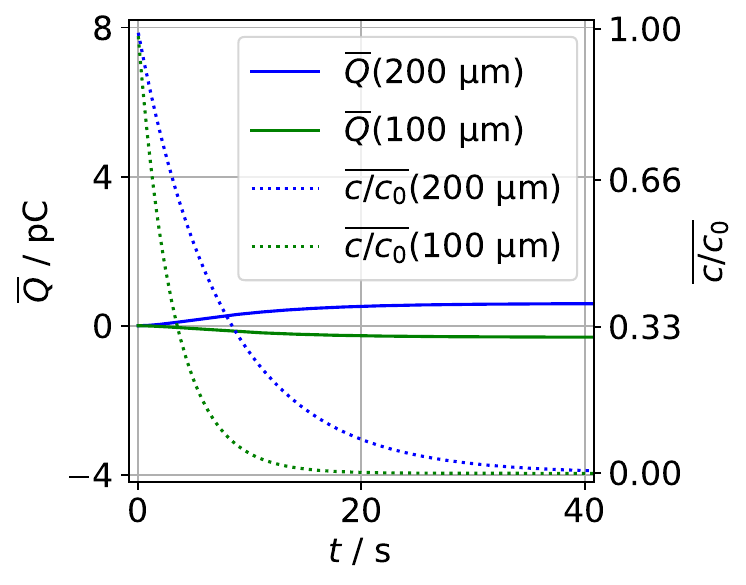}\label{fig:q-mu-ps200100}}\quad
\subfloat[PMMA, bidisperse]{\includegraphics[trim=0mm 0mm 0mm 0mm,clip=true,height=0.22\textwidth]{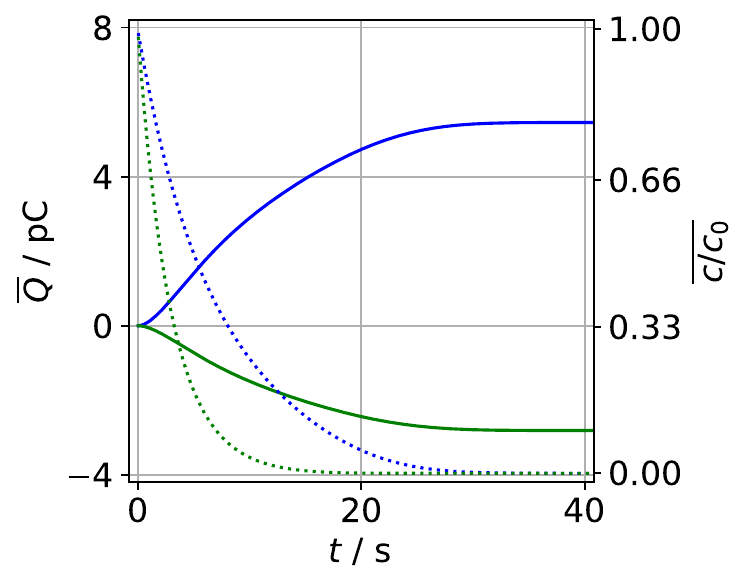}\label{fig:q-mu-pmma200100}}\quad
\subfloat[PS, bidisperse]{\includegraphics[trim=0mm 0mm 0mm 0mm,clip=true,height=0.22\textwidth]{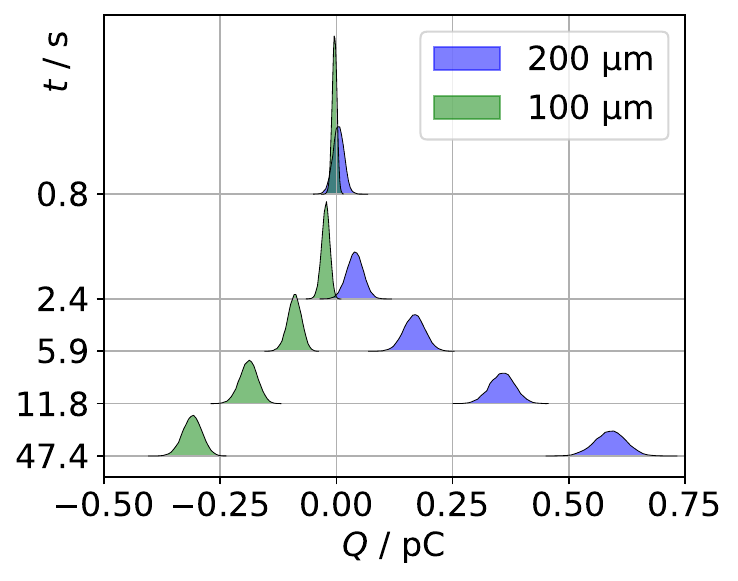}\label{fig:distribution-ps200100}}
\end{center}
\caption[]{%
Size-dependent bipolar charging of bidisperse (a) PS and (b) PMMA particles, as observed in experiments (cf.~\cref{fig:condenserd}).
(c) Charge distributions of PS particles, separated by size: 100~µm (green) and 200~µm (blue).
These simulations consider particle-particle charging in a periodical box without fluid forces.}
\label{fig:bipolar}
\end{figure*}

Third, \cref{fig:bipolar} shows the capability of the SSM to predict \emph{size-dependent bipolar charging}.
For this investigation, we consider only charge transfer between particles and do not account for charge transfer between particles and walls.
Thus, we entirely omit the effect of walls and wall-bounded turbulence.
Instead, the particles are released in a cubical box of 5~cm side length with an initial velocity of 5~m/s (simulations~5 and~6; cf.~\cref{tab:cfd}).

For both particle materials, PS and PMMA, the distributions in \cref{fig:bipolar} show size-dependent bipolar charging.
The underlying mechanism is asymmetric charging, that is, different depletion rates of $c/c_0$ on the surfaces of smaller and larger particles.
The active charging sites on the smaller particles deplete more quickly.
Thus, starting from the second contact, the average net charge tends to transfer from the larger to the smaller particles.

In our case with charging Types~A1 and~B1 (\cref{fig:concept2}), i.e., $\epsilon_p < 0$, this results in a net negative charge transfer from the larger to the smaller particles.
However, the charge transfer during each particle-particle contact is stochastic.
Therefore, while larger particles tend to acquire a net positive charge on average, some large particles may locally and temporarily charge negatively.
This reversal is evident in \cref{fig:distribution-ps200100}, where, up to $t=3$~s, the charge distributions of small PS particles extend into the positive domain, and the distributions of large PS particles extend into the negative domain.

If the reference impact reveals another charging type, such as A2, B2, or C2, then the small particles would charge positively and the large ones negatively instead.
Hence, we do not impose any assumptions on the charge carrier properties that would predetermine the polarity.
The particle polarity emerges from the statistical parameters of the reference impact.

Interestingly, the saturation charge of the particles depends not only on their size but, in contrast to the simulations shown in \cref{fig:q-c-t-200}, also on the particle material.
Here, the saturation of particle charge is not governed by the charge transfer from the wall to the particle and the associated decline of $\alpha_w/\alpha_0$.
Instead, in the periodical box simulations (\cref{fig:bipolar}), charging terminates when all charging sites on all particles are depleted.
Because the particles follow random collision histories and thus accumulate charge at different rates, the final charge is not uniform across particles (as in \cref{fig:distribution-ps200,fig:distribution-pmma200}) but follows a distribution.
The final charge distribution for PS particles at $t=47.4$~s is shown in \cref{fig:distribution-ps200100}.

To summarize the CFD simulations, the SSM successfully predicts the three particle charging patterns highlighted in the Introduction:
\emph{variable impact charge}, \emph{charge reversal}, and \emph{size-dependent bipolar charging}, all of which are observed in experiments.
These behaviors are not the result of manual tuning or selective parameter choices.
They arise naturally from the model, based solely on the scaling laws and the statistical parameters obtained from the reference impact.

\section{Discussion}
\label{sec:discussion}

In this section, we discuss the accuracy and physical foundations of the SSM.

\subsection{Physical assumptions of the SSM}

Compared to other existing particle charging models, the SSM, in particular \cref{eq:muw-scale2,eq:mui-scale2,eq:sigmai-scale,eq:gammai-scale,eq:N}, relies on few simplifications of the charge transfer physics, which are listed and discussed in the following.

A central simplification (\cref{fig:concept1}) is the separation of the impact charge between two surfaces into two independent charge transfers in opposite directions:
A deterministic transfer from a conductor and a stochastic transfer from an insulator.
The experiments of \citet{Mat03} support this assumption.
They showed that the impact charge of a conductive particle on a conductive target varies very little, whereas the impact charge from an insulating particle exhibits strong variability.
The assumption that the charge-transfer distribution for same-material impacts can be derived from the stochastic component of the different-material distributions is, for the PMMA particles under consideration, validated in \cref{fig:binom-3}.

The current formulation models a single stochastic process on the surface of insulators:
The depletion of charging sites, which is equivalent to the depletion of transferable charge carriers.
One could imagine multiple overlapping stochastic processes, such as layers of different charging sites with distinct statistical parameters.
Such processes could be included by introducing additional statistical parameter sets and scaling laws.
However, linking the combined statistics of multiple concurrent stochastic processes to a single reference impact would be challenging.
Likewise, nanoscale morphological changes due to repeated impacts that alter the impact statistics~\citep{Sob25} would require a new conceptual framework.

The model further assumes that $c_0$, $\epsilon_w$, and $\epsilon_p$ are constant for a given material.
This assumption aligns with the findings of \citet{Grosj23}, who observed that the ratio between the correlation length of mosaic features and the donor/acceptor site size remains approximately constant across a material system.
Nonetheless, this assumption demands a dedicated reference impact experiment for each powder under investigation, since the surfaces of even nominally identical powders may differ due to specific manufacturing, handling, or storage conditions.

As stated above, these physical simplifications are few compared to other charging models.
Thus, the SSM serves as a flexible mathematical framework that can, via the input parameters $\mu_0$, $\sigma_0$, $\gamma_0$, and $\Delta Q_{0,\mathrm{min}}$, predict a wide range of particle charging patterns.

As a stochastic model, the SSM does not aim to predict the charge evolution of individual particles but rather the evolution of the charge distributions of the entire powder.
Accordingly, it does not attempt to resolve mechanisms that depend on the history of individual particles~\citep{Grosj23a} but captures powder-wide charge distributions resulting from such mechanisms' aggregation.
We use a different particle in each of the total 869 reference impact experiments to capture these global effects.

This statistical treatment is consistent with the turbulence and particle dynamics modeling in \cref{sec:cfd}, which is also meaningful only in a statistical sense.
In other words, a hypothetical deterministic model that predicts the charge evolution of every individual particle would not improve the CFD simulation's overall accuracy.

The detailed modeling of the individual terms on the right-hand side of \cref{eq:N} is not unique to the SSM but borrowed from contact mechanics and other disciplines.
Therefore, we do not discuss their physical assumptions here.
It is worth emphasizing, however, that the SSM expresses these terms as ratios rather than absolute values.
For instance, the absolute values of the charging site densities $c$ and $c_0$ are not experimentally accessible and must typically be estimated.
In contrast, the ratio $c/c_0$ is much easier to model, particularly since $c/c_0=1$ initially by definition, even though $c$ is unknown.

Similarly, modeling contact areas via a Hertzian approach involves several physical assumptions that are clearly violated for real particles.
Surface roughness, for example, can reduce the effective contact area by several orders of magnitude compared to the Hertzian prediction~\citep{Jan24a}.
However, in the ratios $A/A_0$ (\cref{eq:aw,eq:aij}), many of these assumptions, including all material parameters, cancel out.
Since surface roughness affects both $A$ and $A_0$ equally, the relative error in their ratio is substantially smaller than the error in their absolute values.

Therefore, using ratios instead of absolute values is powerful and significantly reduces the uncertainties associated with the terms in \cref{eq:N}.

\subsection{Error analysis}
\label{sec:err}

\begin{figure}[t]
\begin{center}
\subfloat[Average charge evolution.]{
\begin{minipage}[t]{0.47\textwidth}
\vspace{-\ht\strutbox}
\centering
\includegraphics[trim=0mm 0mm 0mm 0mm,clip=true,width=0.47\textwidth]{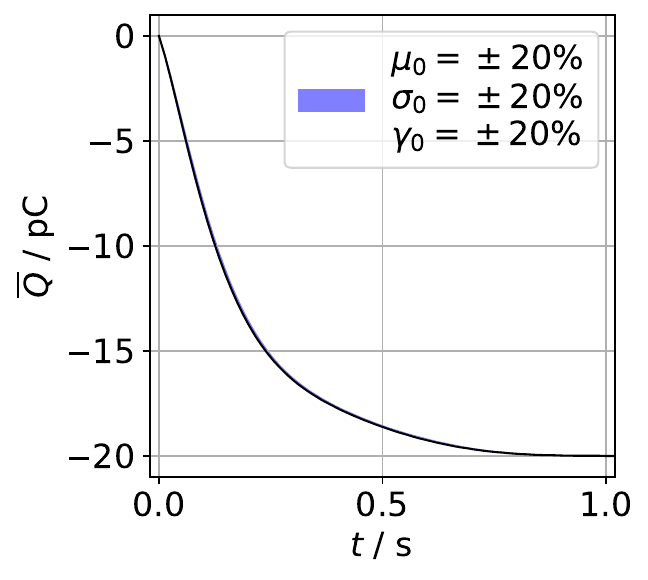}~~
\includegraphics[trim=0mm 0mm 0mm 0mm,clip=true,width=0.43\textwidth]{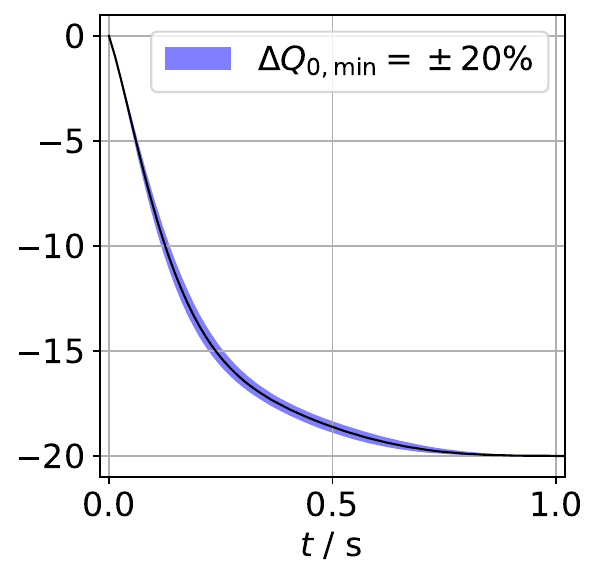}
\end{minipage}
\label{fig:err-t}
}\\
\subfloat[Charge distribution at $t=0.1$~s.]{
\begin{minipage}[t]{0.47\textwidth}
\vspace{-\ht\strutbox}
\centering
\includegraphics[trim=0mm 0mm 0mm 0mm,clip=true,width=0.47\textwidth]{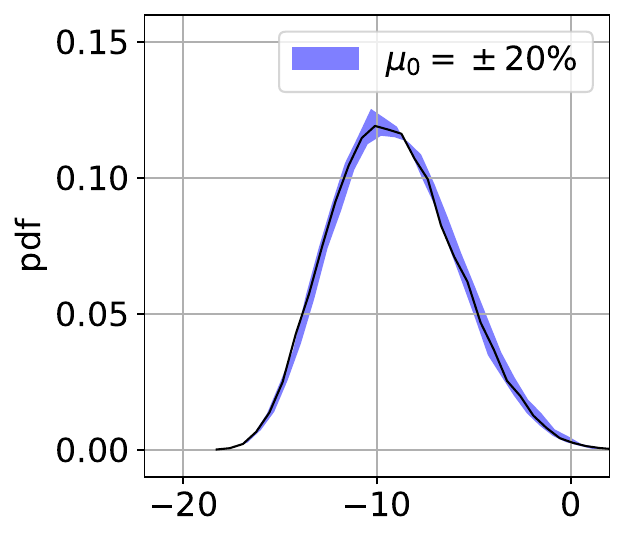}~~
\includegraphics[trim=0mm 0mm 0mm 0mm,clip=true,width=0.44\textwidth]{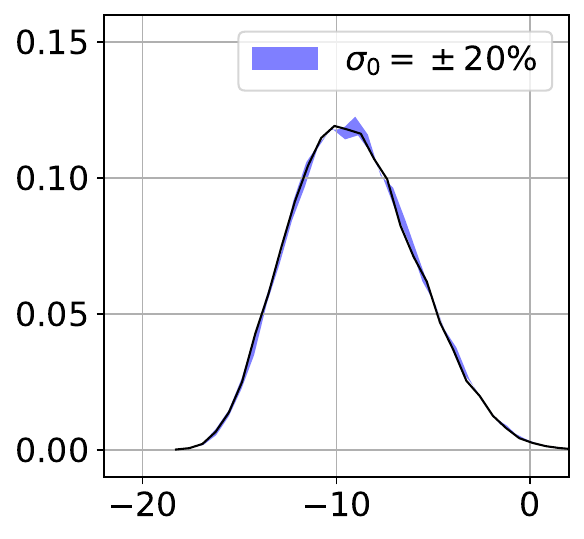}\\
\includegraphics[trim=0mm 0mm 0mm 0mm,clip=true,width=0.47\textwidth]{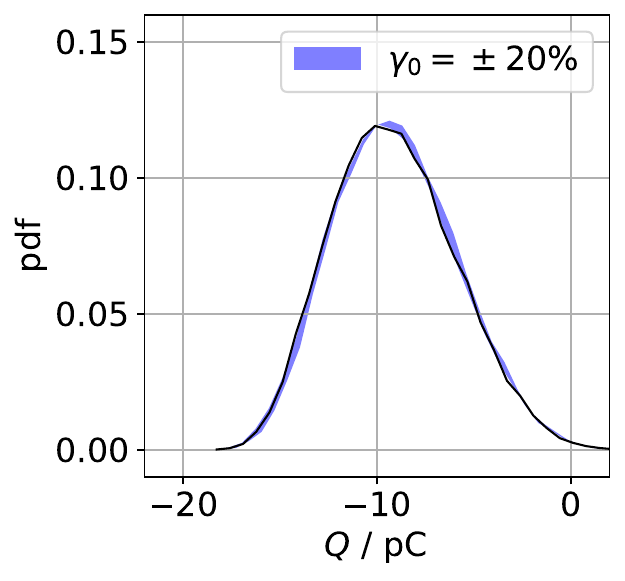}~~
\includegraphics[trim=0mm 0mm 0mm 0mm,clip=true,width=0.44\textwidth]{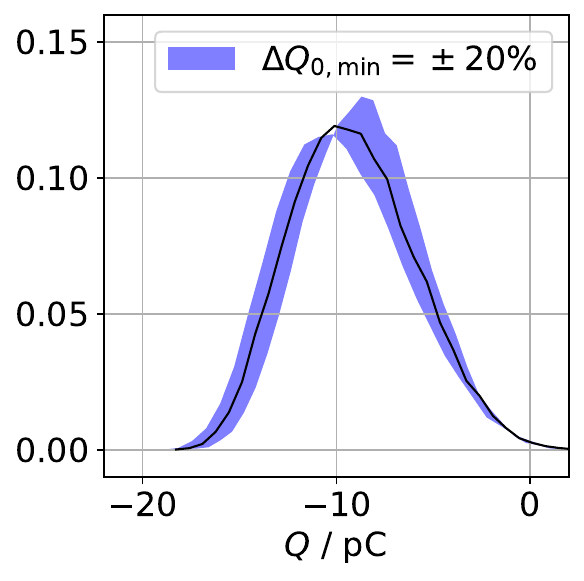}
\end{minipage}
\label{fig:err-distribution}
}
\end{center}
\caption{%
The black curves show (a) the evolution of the average charge and (b) the charge distribution at $t=0.1$~s of 200~µm sized PMMA particles.
The blue shaded areas envelop their variations resulting from a $\pm 20\%$ change in one of the four SSM input parameters.
The simulations include particle-particle and particle-wall charging in a cubical box without fluid forces.}
\label{fig:err}
\end{figure}

In \cref{sec:reference-impact}, we evaluated the statistical uncertainty of the SSM’s four input parameters, $\mu_0$, $\sigma_0$, $\gamma_0$, and $\Delta Q_{0,\mathrm{min}}$, along with the associated error introduced through scaling from one impact to another. 
These uncertainties typically ranged between 10\% and 30\%.

To assess the effect of these uncertainties on model predictions, \cref{fig:err} illustrates the resulting variations in particle charge distributions caused by deliberate changes to the input parameters. 
For this analysis, we employed the same cubical box setup without fluid forces as in simulations~5 and~6, but with wall boundary conditions instead of periodic ones (simulations 7--13; cf.~\cref{tab:cfd}).

We first simulated the charging of PMMA particles using the nominal reference impact parameters listed in \cref{tab:reference-impact}.
Then, we varied each input parameter independently by +20\% and -20\% to quantify its influence. 
This uniform variation allows for a direct comparison of parameter sensitivity.

For the evolution of the average particle charge (\cref{fig:err-t}), variations in $\Delta Q_{0,\mathrm{min}}$ had the largest effect, while changes in $\mu_0$, $\sigma_0$, and $\gamma_0$ had only minor influence.
Similarly, for the particle charge distribution at $t = 0.1$~s (\cref{fig:err-distribution}), $\Delta Q_{0,\mathrm{min}}$ again showed the strongest impact, followed by $\mu_0$.
In contrast, variations in $\sigma_0$ and $\gamma_0$ led to only modest changes in the distribution.

In summary, the model is relatively insensitive to uncertainties in $\sigma_0$.
While $\mu_0$ strongly influences the results, it is also the most accurately determined parameter from the reference impact experiments (see \cref{sec:reference-impact}).
Conversely, $\gamma_0$ is the least precisely determined input, but its influence on model output is minimal.

The parameter $\Delta Q_{0,\mathrm{min}}$ poses particular challenges.
A source of uncertainty is the classification of the impact type, i.e., whether it falls into Type A1, B1, or C1, or alternatively A2, B2, or C2 (cf.~\cref{fig:concept2}). 
Assuming that the transferable charge carriers on conductors are electrons favors Types A1 and B2;
however, the possibility of Type A2 cannot be entirely ruled out.

Further, unlike the other statistical parameters, $\Delta Q_{0,\mathrm{min}}$ does not emerge from a converging distribution but is instead linked to a single extremal event in the experimental dataset. 
Although the convergence curve shown in \cref{fig:reference-impact-convergence} suggests that $\Delta Q_{0,\mathrm{min}}$ can be estimated with relatively few experiments, we overall consider it the least reliable component of the SSM.

\subsection{Relation to other models}


The variations of the statistical parameters plotted in \cref{fig:spcc,fig:spcd} result in the SSM to predict a large range of possible impact outcomes.
The popular condenser model, discussed in the Introduction, is a special case of the SSM.
More specifically, the SSM recovers the condenser model for $c/c_0=0$, which means the charge transfer is deterministic, and $\Delta Q_{0,\mathrm{min}} = -CU$, where $C$ is the electric capacity and $U$ the contact potential difference between the particle and the wall.

Also, the surface state model is a special case of the SSM.
The SSM recovers the surface state model for particle-particle impact charging by setting $\sigma_0$ to zero.
Then, the charge transfer between particles becomes deterministic.

Thus, the SSM is a unified model that can predict same- and different-material particle charging.
The SSM does not contradict existing particle-scale models;
they are special cases of the SSM, which specific input parameters can recover.

Since the presented SSM builds on the mosaic model of \citet{Grosj23}, the particle-scale quantities of the SSM can be traced back to many of the nanoscopic quantities of the mosaic model:
The contact area of the SSM, $A$, is the square of the side length, $L^2$, of the mosaic model's quadratic surface.
The charge of one charging site, $\epsilon_p$ and $\epsilon_w$, correlates in the mosaic model to $e(l/l_0)^2$, where $l_0$ is the size of one site that can transfer an elementary charge $e$, and $l$ is the correlation length between charging sites.
The charging site density in the SSM, $c$, is $1/l^2$ in the mosaic model.

The SSM concept of charging sites that are active and charging sites that are inactive (i.e., $\alpha$, green and brown spots in \cref{fig:concept1b}) is similar to the donor probability $p$ in the mosaic model.
The surface density of active charging sites, $\alpha c$, relates to $p/l^2$ in the mosaic model.
The probability of an active charging site to transfer charge, $p$ in the SSM, is denoted by $\alpha$ in the mosaic model.

Further, the mosaic model predicts average- or variance-driven charging regimes, depending on whether the probability of donors on both contacting surfaces is different or equal.
The SSM recovers the same behavior for $\mu_0=0$, where the charge exchange is variance-driven, and for $\mu_0 \ne 0$, where the charge exchange is average-driven.

A difference between the two models is that the mosaic model tracks the availability of donors and acceptors on both contacting surfaces.
The SSM tracks charging sites, which have a similar concept as donors, but assumes an acceptor probability of unity, which is one assumption underlying the surface state model.
Compared to the mosaic model, the SSM applies a generic idea of charging sites without the need to define their physical size or geometry.
Neither does the SSM require an assumption on the type of charge carrier; it can describe both electron and ion transfer.

An important difference between the models is that the mosaic model relies on nanoscopic quantities that are difficult to access experimentally.
All inputs to the SSM are parameters on the particle scale, do not need to resolve the particle surface, and can be measured by one experimental apparatus.

The pure statistical charging model, discussed in the Introduction, differs from the SSM in the number of conditions that need to be measured experimentally.
The statistical model requires measuring the effect of all relevant impact parameters, their variations, and cross-correlations.
Given the large number of impact parameters that affect a particle’s impact charge, a purely statistical model requires an exuberant number of experiments to quantify the complete parameter space.
In contrast, the SSM requires measuring exactly one reference impact condition per powder and scales the resulting parameters to other conditions.

While the statistical model requires parameter variations, the input parameters to the SSM become more accurate when the single reference condition is met in each particle impact experiment.
Therefore, we control the initial and boundary conditions of the particle impact experiments to a high degree (see \cref{sec:reference-impact}).

The essence of the SSM lies in the scaling laws derived in \cref{sec:scaling}.
Even though we decided in \cref{sec:N,sec:A,sec:alpha,sec:c} on specific formulations to estimate the active charging site ratios of \cref{eq:N}, these specific formulations are not inherent to the model.
They can be exchanged for any other expressions, for example, to include more accurate contact mechanics, other contact modes, or any other parameter.

In the above formulation of the SSM, we consider the effects of impact velocity, particle size, and initial charge on the impact charge distributions.
By applying appropriate correlations to the ratios $N_w/N_0$ and $N_i/N_0$, the model can be straightforwardly extended to include other contact modes, impact angles, particle shapes, surface roughness, temperature, humidity, or pressure.

\section{Conclusions}

We have presented the stochastic scaling model (SSM) for particle charging, a unified framework that can predict charging between the same and different materials.
This unification addresses a long-standing challenge in powder flow electrification modeling, allowing a consistent description of charge transfer between insulating particles, conductive walls, and particles in between.
The model's equation system is closed and tested by statistical parameters obtained from a single, highly-controlled reference impact experiment.
Through a set of physically motivated laws, the model scales these parameters to a wide range of particle contacts, eliminating the need for nanoscopic surface resolution.
We designed the reference experiment to ensure maximal control over the initial and impact conditions, enabling a direct test of the scaling laws without any adjustment of model constants.

Our large-scale CFD simulations of wall-bounded turbulent flows involve 300\,000 insulating particles undergoing millions of collisions with each other and the conductive walls.
Even though the computed particle surface area is 400~million times larger than in the earlier mosaic model, the SSM required less than 0.01\% of the total simulation time.
The model reproduces three cornerstone experimental observations of triboelectric charging:
Variable impact charge even for identical impact conditions, charge reversal, and size-dependent bipolar charging.
By bridging nanoscale physics and system-scale simulations with minimal experimental input, the SSM offers a shift from surface-resolved or empirical models to a versatile statistical framework.

In the future, the model would be further improved by closing the equation system without $\Delta Q_{0,\mathrm{min}}$, the largest source of uncertainty.
The SSM's concept predestines it for including additional physical effects such as different contact modes, particle shape, surface roughness, temperature, humidity, and pressure.
As a computationally efficient and physically grounded tool, the model enables the simulation of charged particle dynamics across many scientific and engineering disciplines, including volcanic plume electrification, dust storms, pharmaceutical powder processing, and industrial process safety.

\begin{table*}[tb]
\caption{\label{tab:cfd} Parameters of all CFD simulations.}
\begin{ruledtabular}
\begin{tabular}{llllllll}
  & Material & $r$ ($\upmu$m) & $\rho_\mathrm{p}$ (kg/m$^3$) & pnd (\#/m$^3$)      & Flow type & Charging models & Figs. \\
\hline 
1     & PS   & 200 & 1000 & $5.0 \times 10^{8}$ & wall-bounded turbulence  & particle-wall \& particle-particle & 8, 9(a), 9(b), 10\\
2     & PMMA & 200 & 1150 & $5.0 \times 10^{8}$ & wall-bounded turbulence  & particle-wall \& particle-particle & 9(a), 9(c), 14\\
3     & PS   & 100 & 1000 & $5.0 \times 10^{8}$ & wall-bounded turbulence  & particle-wall     & 11(a) \\
4     & PMMA & 100 & 1150 & $5.0 \times 10^{8}$ & wall-bounded turbulence  & particle-wall     & 11(b) \\
5     & PS   & 200 & 1000 & $2.5 \times 10^{8}$ & periodical box, no fluid & particle-particle & 12(a), 12(c) \\
      &      & 100 & 1000 & $2.5 \times 10^{8}$ & \\
6     & PMMA & 200 & 1150 & $2.5 \times 10^{8}$ & periodical box, no fluid & particle-particle & 12(b) \\
      &      & 100 & 1150 & $2.5 \times 10^{8}$ & \\
7--13 & PMMA & 200 & 1150 & $5.0 \times 10^{8}$ & cubical box, no fluid    & particle-wall \& particle-particle & 13\\
\end{tabular}
\end{ruledtabular}
\end{table*}

\appendix
\section{Scaling of $Q_{i,\mathrm{max}}$}
\label{sec:qimax}

During an impact on another surface, the charge transfer from particle $i$ reaches its maximum if all active charging sites simultaneously transfer charge, which is according to \cref{eq:mui}
\begin{equation}
\label{eq:vi}
Q_{i,\mathrm{max}} = N_i \epsilon_p = \dfrac{\mu_i}{p} \, .
\end{equation}
The unknown $p$ can be derived from \cref{eq:mui,eq:sigmai,eq:gammai}.
Combining \cref{eq:sigmai,eq:gammai} leads to
\begin{equation}
\label{eq:iv}
\gamma_i=\dfrac{\epsilon_p(1-2p)}{\sigma_i} \, ,
\end{equation}
and combining \cref{eq:sigmai,eq:mui} leads to
\begin{equation}
\label{eq:v}
\epsilon_p=\dfrac{\sigma_i^2}{\mu_i(1-p)} \, .
\end{equation}
Then, substituting \cref{eq:v} into \cref{eq:iv} yields
\begin{equation}
\label{eq:gamma-alt}
\gamma_i=\dfrac{\sigma_i(1-2p)}{\mu_i(1-p)} \, .
\end{equation}
Solving this equation for $p$, and substituting into \cref{eq:vi} results in the closed expression for $Q_{i,\mathrm{max}}$,
\begin{equation}
Q_{i,\mathrm{max}} = \dfrac{\mu_i \left( \gamma_i \mu_i - 2 \sigma_i \right)}{\gamma_i \mu_i - \sigma_i} \, ,
\end{equation}
which is the upper bound to the skew normal distribution in \cref{eq:skew-normal}.

\section{Mathematical model and simulation set up}
\label{sec:math}

\begin{figure*}[tb]
\begin{center}
\subfloat[]{\includegraphics[trim=0mm 0mm 0mm 0mm,clip=true,height=.22\textwidth]{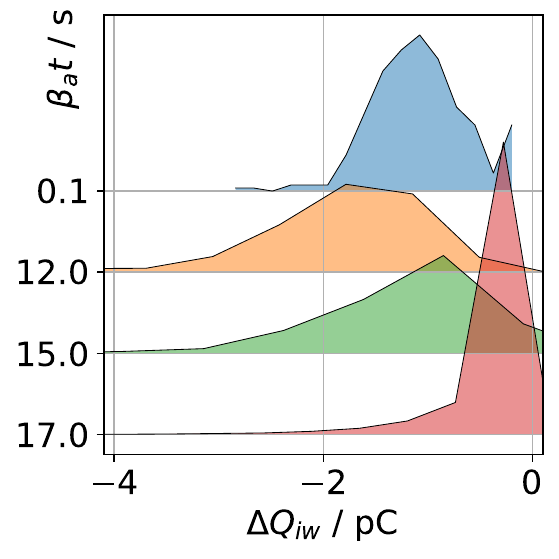}\label{fig:dqpw-distribution-pmma200}}\qquad
\subfloat[]{\includegraphics[trim=0mm 0mm 0mm 0mm,clip=true,height=.22\textwidth]{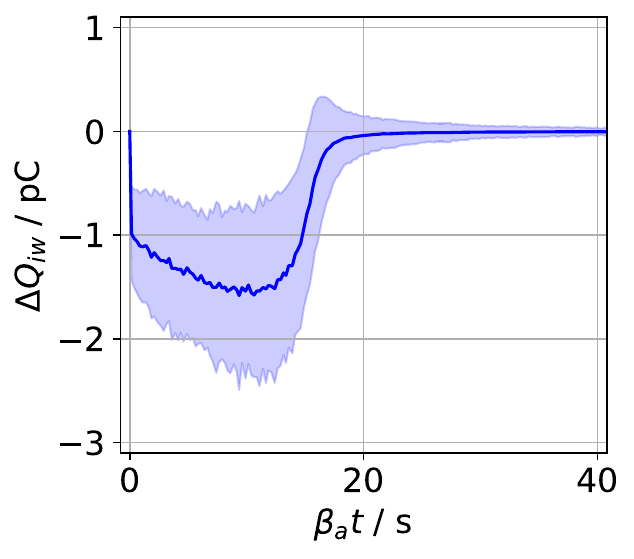}\label{fig:dqpw-t-pmma200}}\qquad
\subfloat[]{\includegraphics[trim=0mm 0mm 0mm 0mm,clip=true,height=.22\textwidth]{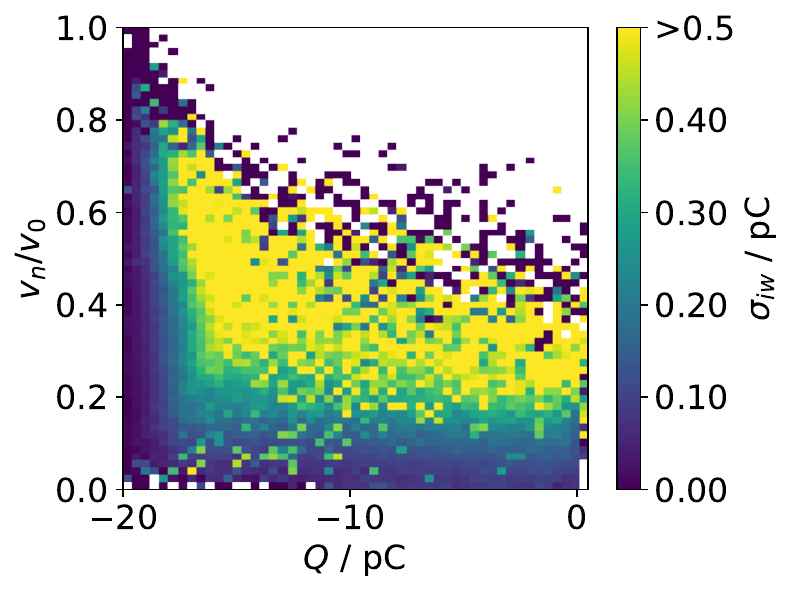}\label{fig:dqpw-heat-pmma200}}
\end{center}
\caption[]{%
Variable impact charge for 200~µm sized PMAA particles, predicted by the SSM, analogous to the impact charge of the PS particles (cf.~\cref{fig:variable-impact-charge-ps})
(a) Charge distributions for selected time instances.
(b) Variation of the impact charge and its standard deviation with time.
(c) Variation of the impact charge for identical impact velocity and initial charge.}
\label{fig:variable-impact-charge-pmma}
\end{figure*}

We implemented the SSM (\cref{sec:spc}) in the open source CFD tool \citet{pafiX},
which employs a coupled Eulerian-Lagrangian approach.
It solves the Navier-Stokes equations for the carrier gas flow in Eulerian framework and tracks each particle individually in Lagrangian framework.
The resulting fluid velocity profiles, particle concentration profiles and the particles' response to the electric field have been extensively validated and documented \citep{Ozl22a,Gro20d,Gro17e} in the last years.


The Navier–Stokes equations for constant densities and diffusivities read
\begin{subequations}
\begin{equation}
\label{eq:mass}
\nabla \cdot \bm{u}_\mathrm{f} =0
\end{equation}
\begin{equation}
\label{eq:mom}
\frac{\partial {\bm u}_\mathrm{f}}{\partial t} + ({\bm u}_\mathrm{f} \cdot \nabla) {\bm u}_\mathrm{f}
= - \frac{1}{\rho_\mathrm{f}} \nabla p_\mathrm{f}  + \nu_\mathrm{f} \nabla^2 {\bm u}_\mathrm{f} + {\bm f}_\mathrm{f} ~.
\end{equation}
\end{subequations}
Therein, ${\bm u}_\mathrm{f}$ is the fluid velocity, $\rho_\mathrm{f}$ the density, $p_\mathrm{f}$ the dynamic pressure, and $\nu_\mathrm{f}$ the kinematic viscosity.

For the simulations presented in \cref{sec:cfd}, we model a section of the duct, apply periodic boundary conditions in the streamwise direction, and drive the flow by the pressure gradient, ${\bm f}_\mathrm{f}$.
In the above equations, second-order schemes approximate the spatial and temporal derivatives.

The computational domain resembles a square-shaped duct with a side length of 5~cm and a streamwise length of 24~cm.
Based on the duct's side length, the flow’s bulk Reynolds number is approximately 5400, and its frictional Reynolds number is 360.
The domain is discretized by 256$\times$144$\times$144 grid points, and turbulence is resolved via direct numerical simulation (DNS).

For each particle of mass $m$, we solve the acceleration, ${\bm a}$, based on Newton’s second law of motion,
\begin{equation}
\label{eq:newton}
m {\bm a} = {\bm F}_\mathrm{d} + {\bm F}_\mathrm{l} + {\bm F}_\mathrm{c} + {\bm F}_\mathrm{e} + {\bm F}_\mathrm{g} ~.
\end{equation}
The drag force acting on the particle, ${\bm F}_\mathrm{d}$, is computed using the empirical correlation of \citet{Put61}.
The lift force, ${\bm F}_\mathrm{l}$, incorporates the formulation of \citet{Saf65} and its correction by \citet{Mei92} to account for particle Reynolds numbers exceeding the shear Reynolds number.
The collisional force, ${\bm F}_\mathrm{c}$, includes both particle-particle and wall-particle interactions.
To detect particle-particle collisions, we implemented ray casting \citep{Roth82,Schr01}, a variant of the hard-sphere collision model.

Electrostatic forces, ${\bm F}_\mathrm{e}$, act on a particle that carries a charge and that is exposed to an electric field.
The electric field at the location of the particle is computed using a hybrid method \citep{Gro17e}.
This approach combines the near-field of the neighbor particle obtained from Coulomb's law with the far-field calculated via Gauss’s law.
The final term in \cref{eq:newton}, ${\bm F}_\mathrm{g}=m{\bm g}$ with ${\bm g}=(9.81,0,0)^\mathrm{T} \mathrm{m\,s^{-2}}$, denotes the gravitational force, indicating that we simulate a downward flow.

\Cref{eq:newton} is numerically integrated using a second-order Crank-Nicolson scheme.

All simulations' parameters are summarized in \cref{tab:cfd}.

\section{Variable impact charge PMMA}
\label{sec:variable-impact-charge-pmma}

For completeness, \cref{fig:variable-impact-charge-pmma} presents the variable impact charge of PMMA particles, analogous to \cref{fig:variable-impact-charge-ps} for PS.
\Cref{fig:variable-impact-charge-ps,fig:variable-impact-charge-pmma} show qualitatively similar behavior.

However, a comparison between the figures reveals that the statistical parameters of the charge distributions depend not only on the contact type and temporal evolution but also strongly on the particle material.
The observed temporal change arises from the evolving particle properties, such as the average increase in particle charge and the depletion of charging sites on the particle surfaces (\cref{fig:sim-overview}).
These evolving properties influence the statistical parameters through the ratio $N_i/N_0$ (cf.~\cref{eq:N}).

Overall, PMMA particles have higher impact charges compared to PS.
Consequently, PMMA particles accumulate charge faster.
As $\alpha_w/\alpha_0$ increases quickly, the charge transfer becomes self-limiting already after~20~s.


\nomenclature[]{$\chi_w$}{elasticity parameter wall (m$^2$ N$^{-1}$)}
\nomenclature[]{$\chi_p$}{elasticity parameter particle (m$^2$ N$^{-1}$)}
\nomenclature[]{$Q$}{particle charge (C)}
\nomenclature[]{$Q_\mathrm{sat}$}{saturation of $Q$ (C)}
\nomenclature[]{$\Delta Q$}{impact charge (C)}
\nomenclature[]{$\Delta Q_{ij}$}{$\Delta Q$ of particle $i$ after contact with particle $j$ (C)}
\nomenclature[]{$\Delta Q_{iw}$}{$\Delta Q$ of particle $i$ after wall contact (C)}
\nomenclature[]{$Q_i$}{charge transfer of particle $i$ to other surface (C)}
\nomenclature[]{$Q_{i,\mathrm{max}}$}{maximum $Q_i$ to other surface (C)}
\nomenclature[]{$Q_w$}{charge transfer from wall to particle (C)}
\nomenclature[]{$\mu_0$}{mean of $\Delta Q_{iw}$ for reference impact (C)}
\nomenclature[]{$\sigma_0$}{standard deviation of $\Delta Q_{iw}$ for reference impact (C)}
\nomenclature[]{$\gamma_0$}{skewness of $\Delta Q_{iw}$ for reference impact}
\nomenclature[]{$\Delta Q_{0,\mathrm{min}}$}{minimum $\Delta Q$ of reference impact}
\nomenclature[]{$\mu_w$}{mean of $Q_w$ (C)}
\nomenclature[]{$\mu_i$}{mean of $Q_i$ (C)}
\nomenclature[]{$\sigma_i$}{standard deviation of $Q_i$}
\nomenclature[]{$\gamma_i$}{skewness of $Q_i$}
\nomenclature[]{$\mu_{w0}$}{mean of $Q_w$ for reference impact (C)}
\nomenclature[]{$\mu_{i0}$}{mean of $Q_i$ for reference impact (C)}
\nomenclature[]{$\sigma_{i0}$}{standard deviation of $Q_i$ for reference impact (C)}
\nomenclature[]{$\gamma_{i0}$}{skewness of $Q_i$ for reference impact (C)}
\nomenclature[]{$\epsilon_w$}{charge of one wall charging site (C)}
\nomenclature[]{$\epsilon_p$}{charge of one particle charging site (C)}
\nomenclature[]{$N$}{number of active charging sites}
\nomenclature[]{$N_w$}{$N$ at contact area of wall}
\nomenclature[]{$N_{w0}$}{$N$ at contact area of wall for reference impact}
\nomenclature[]{$N_i$}{$N$ at contact area of particle}
\nomenclature[]{$N_0$}{$N$ at contact area of particle for reference impact}
\nomenclature[]{$p$}{probability that active site transfers charge}
\nomenclature[]{$M$}{number of reference impact or impacts}
\nomenclature[]{$c$}{surface density of charging sites (m$^{-2}$)}
\nomenclature[]{$c_0$}{$c$ for reference impact (m$^{-2}$)}
\nomenclature[]{$\alpha$}{ratio of active to total number of charging sites}
\nomenclature[]{$\alpha_0$}{$\alpha$ for reference impact}
\nomenclature[]{$\alpha_i$}{$\alpha$ for particle to wall transfer}
\nomenclature[]{$\alpha_w$}{$\alpha$ for wall to particle transfer}
\nomenclature[]{$\alpha_{ij}$}{$\alpha$ for particle to particle transfer}
\nomenclature[]{$A_p$}{particle surface area (m$^2$)}
\nomenclature[]{$A$}{contact area (m$^2$)}
\nomenclature[]{$A_0$}{$A$ for reference impact (m$^2$)}
\nomenclature[]{$A_w$}{$A$ between particle and wall (m$^2$)}
\nomenclature[]{$A_{ij}$}{$A$ between two particles (m$^2$)}
\nomenclature[]{$\rho_p$}{particle density (kg m$^{-3}$)}
\nomenclature[]{$k_e$}{restitution coefficient}
\nomenclature[]{$v_n$}{normal impact velocity of particle on wall (m s$^{-1}$)}
\nomenclature[]{$v_0$}{$v_n$ for reference impact (m s$^{-1}$)}
\nomenclature[]{$v_{ij}$}{relative impact velocity of two particles (m s$^{-1}$)}
\nomenclature[]{$r$}{particle radius (m)}
\nomenclature[]{$r_0$}{$r$ in reference impact (m)}
\nomenclature[]{$\beta_a$}{artificial increase of $A$}
\nomenclature[]{$\beta_c$}{increase of $A$ due to contact mode}
\nomenclature[]{$\beta_r$}{increase of $A$ due to surface roughness}
\nomenclature[]{$E$}{electric field strength at particle surface (V m$^{-1}$)}
\nomenclature[]{$E_i$}{$E$ at particle-wall contact (V m$^{-1}$)}
\nomenclature[]{$E_{ij}$}{$E$ at particle-particle contact (V m$^{-1}$)}
\nomenclature[]{$E_\mathrm{sat}$}{saturation of $E$ (V m$^{-1}$)}
\nomenclature[]{$C_\mathrm{sat}$}{fitting parameter for \cref{eq:qsat} (C m$^{-2}$)}
\nomenclature[]{$\varepsilon$}{electric permittivity of gas phase (C V$^{-1}$ m$^{-1}$)}

\vspace{-\baselineskip}
\renewcommand{\nomname}{}
\setlength{\nomitemsep}{-\parsep}
\section{Nomenclature of the SSM}
\vspace{-13mm}
\printnomenclature

\begin{acknowledgments}
This project has received funding from the European Research Council~(ERC) under the European Union’s Horizon 2020 research and innovation programme~(grant agreement No.~947606 PowFEct) and the Deutsche Forschungsgemeinschaft (DFG, German Research Foundation) -- Projektnummer~562343622.
\end{acknowledgments}

\bibliography{\string~/essentials/publications}

\end{document}